\documentclass[11pt,a4paper]{article}
\pdfoutput=1
\usepackage{jheppub}
\usepackage{amsmath,epsfig}

\usepackage{mathrsfs}
\usepackage{amssymb}
\usepackage{mathtools}
\usepackage{graphics}
\usepackage{axodraw}
\usepackage{slashed}
\usepackage{url}
\usepackage{hyperref}
\usepackage{enumitem}

\usepackage{color}
\usepackage[normalem]{ulem}

\usepackage{multirow}
\usepackage[svgnames,dvipsnames,x11names]{xcolor}
\usepackage[framemethod=default]{mdframed}
\newmdenv[skipabove=7pt,
skipbelow=7pt,
rightline=true,
leftline=true,
topline=true,
bottomline=true,
backgroundcolor=gray!7,
linecolor=gray,
innerleftmargin=5pt,
innerrightmargin=5pt,
innertopmargin=0pt,
innerbottommargin=10pt,
leftmargin=0cm,
rightmargin=0cm,
linewidth=1.5pt]{eBox}

\usepackage{gensymb}

\newcommand{\be}{\begin{equation}}
\newcommand{\ee}{\end{equation}}
\newcommand{\bea}{\begin{eqnarray}}
\newcommand{\eea}{\end{eqnarray}}

\newcommand{\eps}{\epsilon}

\usepackage{color}

\usepackage{fontawesome5}
\definecolor{color_git}{rgb}{0.098, 0.160, 0.345}
\newcommand{\gitlink}{\href{https://github.com/stefanmarinus/amiqs}{\textsc{g}it\textsc{h}ub {\large\color{color_git}\faGithub}}}

\title{Predicting the baryon asymmetry with degenerate right-handed neutrinos }

\preprint{IFIC/23-18, FTUV-23-0519.8654}

\author[]{S. Sandner,}
\author[]{ P. ~Hern\'andez,}
\author[]{J. ~L\'opez-Pav\'on,}
\author[]{and N. Rius}

\affiliation[]{Instituto de F\'{\i}sica Corpuscular, Universidad de Valencia and CSIC, 
 Edificio Institutos Investigaci\'on, Catedr\'atico Jos\'e Beltr\'an 2, 46980 Spain}
 
 \emailAdd{stefan.sandner@ific.uv.es}
 \emailAdd{m.pilar.hernandez@uv.es}
 \emailAdd{jlpavon@ific.uv.es}
\emailAdd{nuria.rius@ific.uv.es}

\abstract{
We consider the generation of a baryon asymmetry in an extension of the Standard Model with two singlet Majorana fermions that are degenerate above the electroweak phase transition. The model can explain neutrino masses as well as the observed matter-antimatter asymmetry, for masses of the heavy singlets below the electroweak scale. The only physical CP violating phases in the model are those in the  PMNS mixing matrix, i.e. the Dirac phase and a Majorana phase that enter light neutrino observables. We present an accurate analytic approximation for the baryon asymmetry in terms of CP flavour invariants, and derive the correlations  with neutrino observables. We demonstrate that the measurement of CP violation in neutrino oscillations as well as the mixings of the heavy neutral leptons with the electron, muon and tau flavours suffice to pin down the matter-antimatter asymmetry from laboratory measurements. 
}


\begin{document}

\maketitle


\section{Introduction}
\label{sec:introduction}

It is well known that the new physics responsible for neutrino masses could also seed the matter-antimatter asymmetry of the Universe~\cite{Fukugita:1986hr}. 
An interesting question is to what extent the asymmetry is connected to the CP violating phases in the neutrino mass matrix, that we can hope to measure in future neutrino experiments. 
Generically, this connection is rather loose, because extensions of the Standard Model that can account for neutrino masses involve CP-violating heavier sectors that contribute to the matter-antimatter asymmetry. 
Therefore, unless we can also test experimentally those sectors and determine the CP-violating phases affecting their dynamics, it is not possible to predict the baryon asymmetry. 

The exception to this conclusion is when the simplicity of the model restricts the number of physical CP-violating phases to be the same as those in the light neutrino mass matrix. 
This can happen when the model has naturally minimal flavour violation~\cite{DAmbrosio:2002vsn,Davidson:2006bd,Gavela:2009cd}, as happens for example in the Type II seesaw model~\cite{LAZARIDES1981287, PhysRevD.23.165, MAGG198061}, or in the presence of flavour symmetries~\cite{Pascoli:2006ie,Pascoli:2006ci,Hagedorn:2016lva}. 
One example  of the latter is the minimal Type I seesaw model~\cite{Minkowski:1977sc, GellMann:1980vs, Yanagida:1979as, Mohapatra:1979ia} with two degenerate Majorana neutrinos. 
The $O(2)$ symmetry of the Majorana mass matrix is broken by the Yukawa couplings, so the heavy neutrinos are degenerate above the electroweak phase transition and get a tiny splitting below the transition. 
It can be shown that this model can account for the light neutrino masses and only has two physical CP-violating phases, that can be parametrized by those in the neutrino mass matrix: a Dirac phase and a Majorana phase. 

The generation of the baryon asymmetry has been considered in the minimal Type I seesaw model in many previous works~\cite{Akhmedov:1998qx,Asaka:2005pn,Canetti:2010aw, Canetti:2012vf, Canetti:2012kh, Shuve:2014zua, Hernandez:2015wna, Abada:2015rta, Hernandez:2016kel, Hambye:2016sby, Drewes:2016jae, Drewes:2016gmt, Ghiglieri:2017csp, Hambye:2017elz, Antusch:2017pkq, Eijima:2018qke, Abada:2018oly, Klaric:2021cpi, Hernandez:2022ivz}, including the degenerate limit \cite{Antusch:2017pkq}.
However, the strong correlations between the baryon asymmetry and CP violating neutrino observables that exist in this limit have not been clarified. The objective of this paper is two fold. 
First, we develop accurate analytical approximations to the baryon asymmetry in the degenerate limit, following the methods introduced in~\cite{Hernandez:2022ivz}. 
This allows us to identify the relevant CP flavour invariants entering the baryon asymmetry, that can then be easily written in terms of  neutrino observables, notably the CP-violating phase participating in neutrino oscillations and the Majorana phase of the light neutrino mass matrix. 
Second, we study the predictivity of the baryon asymmetry in this model from future measurements: flavour mixings and masses of the heavy neutral leptons (HNLs), CP violation in neutrino oscillations and neutrinoless double-beta decay. 

The structure of the paper is as follows.
In Sec.~\ref{sec:model} we introduce the model and discuss the conditions under which a baryon asymmetry can be generated.
This leads to the identification of distinct washout regimes to which CP flavour invariants can be associated.
In Sec.~\ref{sec:cpinv} we derive the leading order CP flavour invariants in terms of observables.
Sec.~\ref{sec:ana_approx} is devoted to the analytical solutions of the Boltzmann equations governing the evolution of the baryon asymmetry and their relation to the CP invariants. 
The analytical results are then applied in Sec.~\ref{sec:parambounds} to derive robust constraints on the model parameter space arising from the observed baryon asymmetry. 
In Sec.~\ref{sec:num} we present a numerical  scan of parameter space and show the validity of the analytical approximations.
We discuss in Sec.~\ref{sec:potential_measure} how measurements of the flavour dependent HNL mixing together with the determination of the Dirac CP phase are enough to predict both, the Majorana phase as well as the baryon asymmetry and compare to the non-degenerate scenario.
We conclude in Sec.~\ref{sec:conclu}.


\section{The model, Sakharov conditions and CP invariants}
\label{sec:model}

 In this work we focus on the minimal Type I seesaw model which can successfully explain all neutrino oscillations constraints.
 It includes two Majorana fermion singlet states $N^{i}$, which we will assume to be \textit{exactly} degenerate before the electroweak phase transition (EWPT).
 The renormalizable Lagrangian takes the form
 \begin{eqnarray}
{\cal L} = {\cal L}_{SM}- \sum_{\alpha,i} \bar L^\alpha Y^{\alpha i} \tilde\Phi N^i - \sum_{i,j=1}^{2} {1\over 2} \bar N^{ic} M_{Rij} N^j+ h.c.\,, 
\label{eq:lag}
\end{eqnarray}
where $Y$ is a $3\times 2$ complex Yukawa interaction matrix and $M_R$ is a $2\times 2$ complex symmetric matrix with two degenerate eigenvalues.
 $L$ represents the fermion doublet and $\tilde\Phi = i\sigma_2 \Phi^*$ is the Higgs doublet with a vacuum expectation value of $\langle \Phi\rangle =v = 246/\sqrt{2} \, \rm{GeV}$.
The  spectrum of this model contains four massive and one massless neutrinos of Majorana nature, out of which the three lighter are the observed neutrinos and the two heavier are referred to as HNLs.
The light neutrino mass matrix is given by the seesaw formula
\be
-m_\nu={v^2}Y M_R^{-1} Y^T=\left(U_\nu^*m\,U_\nu^\dagger\right)_{\alpha\beta}\,,
\label{eq:mnu}
\ee
where $U_\nu(\theta_{12},\theta_{13},\theta_{23},\delta,\phi)$  is the PMNS matrix\footnote{We use the parameterization of the PDG \cite{ParticleDataGroup:2020ssz}.} describing the light neutrino mixing, and $m$ is the diagonal matrix of the light neutrino masses. Notice that in this simple case of degenerate HNL before the EWPT, there are only nine extra parameters beyond the Standard Model ones. 
The  measured light neutrino mass splittings and mixings fix five of them, while the Majorana mass scale, i.e. the eigenvalue of the matrix, $M_R$,  one of the eigenvalues of $Y^\dagger Y$, which generically has two distinct ones, and two complex CP violating phases of the PMNS matrix (one Dirac CP phase and one Majorana phase) are presently unconstrained.

Direct and indirect searches for HNL severely constrain their masses below ${\mathcal O}(100 \rm{MeV})$ range.
The $\rm{GeV}$ range instead is currently poorly explored experimentally,  and future experiments will be sensitive to a very large fraction of 
the available parameter space.
In the naive one family approximation the expected interaction strength of the HNLs with the SM content is expected to be 
\be
\Theta \sim v Y M_R^{-1} \sim {\mathcal O}\left(\sqrt{m_\nu \over M_R}\right)\,.
\label{eq:theta}
\ee
This relation suggests that for $M_R = \mathcal{O}(\rm{GeV})$ the HNLs would be elusive experimentally. 
However, for certain textures of $Y$ and $M_R$ this naive estimate breaks down~\cite{Wyler:1982dd,Mohapatra:1986aw,Mohapatra:1986bd,Bernabeu:1987gr,Branco:1988ex,Akhmedov:1995ip,Barr:2003nn,Kersten:2007vk,Gavela:2009cd}.
In particular, this is the case for models which have an unconventional lepton number (LN) symmetry.
Assigning the LN
\be
\label{eq:LN_assignment}
L(N_1)= -L(N_2)=1\,
\ee
for the HNLs the symmetric texture takes the form \cite{Gavela:2009cd}
\be
Y_{\alpha i}=\begin{pmatrix}
y_{e} & 0\\
y_{\mu} & 0 \\
y_{\tau} & 0 
\end{pmatrix},\;\;
M_R=\begin{pmatrix}
0 & \Lambda \\
\Lambda  & 0 
\end{pmatrix}\, .
\label{eq:LNtexture}
\ee
Such scenarios lead to vanishing neutrinos masses in Eq.~(\ref{eq:mnu}), while $\Theta$ is unsuppressed. 
Note that in this limit one combination of the sterile states does not couple to leptons, since $Y^\dagger Y$ has a vanishing eigenvalue.
As expected from the general LN violating nature of the Majorana neutrinos, the light neutrino masses are directly proportional to the symmetry breaking parameters, $y'_\alpha$:
\be
Y_{\alpha i}=\begin{pmatrix}
y_{e} e^{i \varphi_e} & y'_e e^{i \varphi_e'}\\
y_{\mu} e^{i \varphi_\mu} & y'_{\mu}  e^{i \varphi_\mu'}\\\
y_{\tau} e^{i \varphi_\tau} & y'_{\tau} e^{i \varphi_\tau'}\ 
\end{pmatrix},\;\;
M_R=\begin{pmatrix}
0 & \Lambda \\
\Lambda  & 0 
\end{pmatrix}\,,
\label{eq:LNVparam2N}
\ee
with $y_{\alpha}, y'_{\alpha}, \Lambda\in \mathbb{R}^+$. 
The diagonal entries of $M_R$ would also break the symmetry, but also the degeneracy of the two HNLs.
Our interest in this current study, however, is focused on degenerate HNLs\footnote{The general scenario of non-degenerate HNLs was studied in~\cite{Hernandez:2022ivz}.} and therefore we assume no symmetry breaking in $M_R$.

As mentioned above, to explain neutrino oscillations data not all parameters of Eq.~\eqref{eq:LNVparam2N} are free but are correlated via the seesaw formula of Eq.~\eqref{eq:mnu} which takes the form
\be
- \left( m_{\nu}\right)_{\alpha\beta} = \frac{v^2}{\Lambda}\left( Y_{\alpha 1} Y_{\beta 2} + Y_{\alpha 2} Y_{\beta 1} + \mathcal{O}\left(y'^{3}_\alpha \right)\right)\,.
\label{eq:mnuLN}
\ee

We follow the framework presented in~\cite{Hernandez:2022ivz} to compute analytically the baryon asymmetry generated in this model by perturbing around the symmetric limit, that is via a series expansion in the small symmetry breaking parameters $y'_\alpha$. In the following we will use the 
quantities $y$, $y'$ and $U^2$
\be
\label{eq:y_yp_U2_def}
y^2 \equiv \sum_\alpha y_\alpha^2, \;\;\; y'^2 \equiv \sum_\alpha y_\alpha'^2,\;\;\;U^2\equiv \frac{1}{2}\sum_{\alpha,I}|\Theta_{\alpha I}|^2. 
\ee

Note that we neglect the running of the HNL mass matrix elements. 
A HNL mass splitting is generated above the EWPT by $T$-independent loop corrections and can be estimated to be~\cite{Casas:1999tp,Roy:2010xq,Lopez-Pavon:2012yda}
\begin{align}
\label{eq:one_loop_dm}
\delta M_{\rm{loop}}\approx\frac{M}{4\pi^2}\rho y y' \log\left(\frac{Q_0}{Q}\right)\,, 
\end{align}
where $Q$ is the energy scale and $Q_0$ is the initial scale for which the splitting is assumed to be zero.
$\rho$ is defined in Eq.~\eqref{rhoNH} and Eq.~\eqref{rhoIH} for normal hierarchy (NH) and inverted hierarchy (IH) respectively. 
It will be shown that this loop induced HNL mass splitting is negligible in the baryon asymmetry generation.
Thus, it is sufficient to consider the HNL mass matrix as given in Eq.~\eqref{eq:LNVparam2N}.

\subsection{Sakharov conditions and weak washout regimes}
\label{subsec:sakharov}

Every dynamical generation of a net baryon asymmetry has to fulfil certain criteria.
These conditions were first derived by Sakharov~\cite{Sakharov:1967dj} and lead to three conditions for successful baryogenesis named after him:
\begin{itemize}
\item Baryon number violation.
\item C and CP violation.
\item Out of thermal equilibrium dynamics.
\end{itemize}
All of them are fulfilled in the model of Eq.~\eqref{eq:lag}.
In fact, the baryon number violating process arises purely from SM dynamics at high temperature.
Before the EWPT sphaleron processes in the electroweak gauge field sector efficiently reshuffle lepton and baryon number into each other.
To fulfil the remaining two conditions and to explain the observed baryon asymmetry, the HNL interaction with the SM are indispensable.
In the mass range of interest, $M = \mathcal{O}(\rm{GeV})$, the relevant thermal evolution of the HNLs is of freeze-in type.
The CP-violating Yukawa interactions with the thermal plasma produces the states $N^{i}$ via $L_\alpha H \leftrightarrow N^{i}$, or various different types of $2\leftrightarrow 2$ scatterings, with strength $Y_{\alpha i}$.
Naively, the out of equilibrium condition of Sakharov therefore bounds the strength of the HNL plasma interactions in order to prevent full thermalization before  sphaleron freeze-out. 
This consideration is essential to understand the parameter space compatible with a baryon asymmetry.
\vspace{0.25 cm}
\newline
\noindent
\textbf{Weak washout modes.}
In the cosmological context a state is said to be out of thermal equilibrium if its total interaction strength $\Gamma$ is smaller than the Hubble expansion rate $H_u$.
In a physical picture this definition compares the average mean free path between two interactions against the intrinsic expansion of the metric scale at a given moment in time.  
At the time of interest, before the EWPT, the Universe can be modelled by a relativistic plasma which, via the Einstein field equations, leads to 
\be
\label{eq:hubble}
H_u (T)	= \frac{T^2}{M_{\rm{pl}}^*}\,,
\ee
with the re-scaled Planck mass
\be
M_{\rm{pl}}^* \equiv \sqrt{\frac{45}{4 \pi^3 g_*(T)}} m_{\rm{pl}}\,,
\ee
and $m_{\mathrm{pl}} = 1.22 \times 10^{19}\,\mathrm{GeV}$ is the Planck mass.
We assume the number of thermal relativistic degrees of freedom at temperature $T$ to be $g_*(T)= 106.75$ throughout the evolution, that is, we neglect the HNLs contribution.\footnote{The physical expectation is that the baryon asymmetry is linearly proportional to $T_{\rm{EW}}/H_u(T_{\rm{EW}})$ such that the error induced by neglecting the HNL contribution to $g_*$ is always below $\sim 3\%$. }
No net baryon asymmetry is possible if \textit{all} relevant processes involved in its generation are fast compared to the Hubble expansion rate.
The scale of interest of the baryon number violating processes is the EWPT.
Therefore, we can distinguish various regimes depending on what modes satisfy this condition at the electroweak phase transition, $T_{\rm EW}$.

The first relevant scale, rather than being a thermalization rate, is related to the time at which the CP asymmetries can be generated.
In the non-degenerate case, $M_1 \neq M_2$, usually considered~\cite{Akhmedov:1998qx, Asaka:2005pn}, this scale is given by the HNL vacuum oscillations
\begin{align}
\label{eq:rate_vacuum_osc}
\Gamma_{\mathrm{osc}}^{\mathrm{vac}} \sim \frac{M_2^2 - M_1^2}{T}\,.
\end{align}
In this model, however, the HNLs are degenerate and the oscillation rate depends on thermally induced mass differences, more precisely the relevant scale can be estimated as\footnote{The rate can be estimated as the off-diagonal term of the Hamiltonian, see Eq.~\eqref{eq:H_VM}, in the basis where the helicity conserving HNL interactions are diagonal.}
\begin{align}
\label{eq:rate_thermal_osc}
\Gamma_{\mathrm{osc}}^{\mathrm{thm}} \sim \rho y y' T\,.
\end{align}
The interference between the CP-even oscillation phases  and the CP-odd phases in the helicity conserving interactions lead to a net CP asymmetry. Introducing Eq.~\eqref{eq:one_loop_dm} in Eq.~\eqref{eq:rate_vacuum_osc}, we can compare the loop induced vacuum oscillation rate with the thermal oscillation rate given by the above equation as
\begin{align}
\label{eq:loop_vs_th}
\frac{\Gamma_{\mathrm{osc}}^{\mathrm{loop}}}{\Gamma_{\mathrm{osc}}^{\mathrm{thm}}} \sim 5\times 10^{-2}\left(\frac{M}{T}\right)^2 \log\left(\frac{Q_0}{Q}\right),
\end{align}
and thus the loop induced HNL splitting can be safely neglected in our analysis, since in the regime of interest $M \ll T$.

The first relevant thermalization scale is the one related to the interaction rates of the HNL. At temperatures such that $T\gg M_i$, the HNLs can be assumed to be relativistic in the corresponding processes.  
In this case the rate is given by
\begin{eqnarray}
\Gamma \propto {\rm Tr}[Y Y^\dagger] T.
\end{eqnarray}
It can be shown that the region of parameter space within experimental reach satisfies $\Gamma(T_{\rm EW}) > H_u(T_{\rm EW})$. 
There are however various states with potentially slow thermalization rates that could be reservoirs of the freeze-in CP asymmetries. 

Firstly, a flavour $\alpha$ may not thermalize.
The interaction rate of flavour $\alpha$ is given by:
\begin{eqnarray}
\Gamma_{\alpha}(T) \propto \epsilon_\alpha \Gamma(T),  \;\;\; \epsilon_\alpha \equiv {(Y Y^\dagger)_{\alpha\alpha} \over { \rm Tr}[Y Y^\dagger]}= {y_\alpha^2\over y^2}+\mathcal{O}\left(y'^2/y^2\right)\,.
\label{eq:epsilonalpha}
\end{eqnarray}
A flavour hierarchy in the Yukawa couplings can result in a hierarchy in the corresponding interaction rates and can lead to $\Gamma_\alpha \leq H_u$ at the EWPT, even if $\Gamma > H_u$. 

A second slow mode related to helicity conserving interactions can be found.\footnote{In~\cite{Hernandez:2022ivz} we referred to this regime as weak lepton number violating, since a standard lepton number symmetry, with $\rm{L}(N_1)=  \rm{L}(N_2)=1$, exists in the limit $M\rightarrow 0$. 
To avoid confusion, here we will only refer to LN as that of Eq.~\eqref{eq:LN_assignment} and the limit $M\rightarrow 0$ as helicity violating, following the nomenclature of ref.~\cite{Ghiglieri:2017gjz}.}
The corresponding rate is proportional to (we assume $M\leq T_{EW}$):
\begin{eqnarray}
\Gamma^{\rm slow}_{M} \propto \left({M\over T}\right)^2 \Gamma \leq \Gamma\,.
\label{eq:slowM}
\end{eqnarray}

In contrast with the non-degenerate case considered in ref.~\cite{Hernandez:2022ivz}, even if all the previous rates are fast, there is always one slowly thermalizing mode related to the approximate LN symmetry, Eq.~\eqref{eq:LNtexture}. The symmetry is only broken by the small couplings $y_\alpha'$, that are assumed to be perturbative, and therefore the lepton number thermalization rate is
\begin{eqnarray}
\Gamma^{\rm slow}_{\rm{LN}} \propto y'^2 T\,.  
\label{eq:slowLN}
\end{eqnarray}
Taking into account the constraints from neutrino oscillations, see App.~\ref{app:param}, this can be expressed as
\be
\Gamma^{\rm slow}_{\rm{LN}} \propto y'^2 T \sim \frac{m_\nu^2}{2^3 v^2 U^2} T \,.
\ee
Comparing to Eq.~\eqref{eq:hubble}, it is easy to see that for HNL mixings \textit{larger} than $U^2 \sim 10^{-11}$ this rate does not thermalize by the EWPT, independently of the HNL mass.
Hence, in all relevant parameter space this mode can serve as a reservoir for the baryon asymmetry. 

It is now the task to identify the region of parameter space in which the slow modes of Eqs.~\eqref{eq:epsilonalpha} and~\eqref{eq:slowM} remain out of equilibrium at the EWPT.
The result can be seen in Fig.~\ref{fig:regimes} 
in which we show the different regimes on the plane mixing of the HNL, $U^2$, versus their mass, $M$,  with light neutrino data properly accounted for (see App.~\ref{app:param}). 
For mixings above the dashed-dotted line, helicity conserving interactions are in equilibrium at $T_{\rm EW}$, that is $\Gamma_M \geq H_u$.
Below the line, they are not. 
We refer to the regions above or below the line as {\it strong helicity conservation} (sHC) or {\it weak helicity conservation} (wHC). \\
\noindent
On the other hand, a flavour hierarchy in the Yukawa interactions can lead to
\be
\Gamma_\alpha(T_{\rm EW}) < H_u(T_{\rm EW}) < \Gamma(T_{\rm EW})\,,
\label{eq:flavour}
\ee
for some $\alpha=e,\mu,\tau$. In Fig.~\ref{fig:regimes}, the red dashed lines indicate the band corresponding to Eq.~(\ref{eq:flavour}). 
They depend somewhat on the active neutrino mass hierarchy. For parameters within the band, a flavour direction may remain weakly coupled. We refer to this regime as {\it flavoured}. Note that the band overlaps both with sHC and wHC regimes.

The Sakharov out-of-equilibrium condition is therefore satisfied by one or more modes in the experimentally accessible parameter space indicated by the region inside the black dotted line. 
On the other hand, we explicitly show in App.~\ref{app:unflavoured} that the asymmetry is too small to explain the observed value in all the unflavoured regions. That is, inside the light blue region (above the red dashed band) in Fig.~\ref{fig:regimes}.
Therefore, we can focus on just two regimes:
%
\begin{figure}[!t]
\centering
\begin{tabular}{cc}
\hspace{-0.5cm} \includegraphics[width=0.5\textwidth]{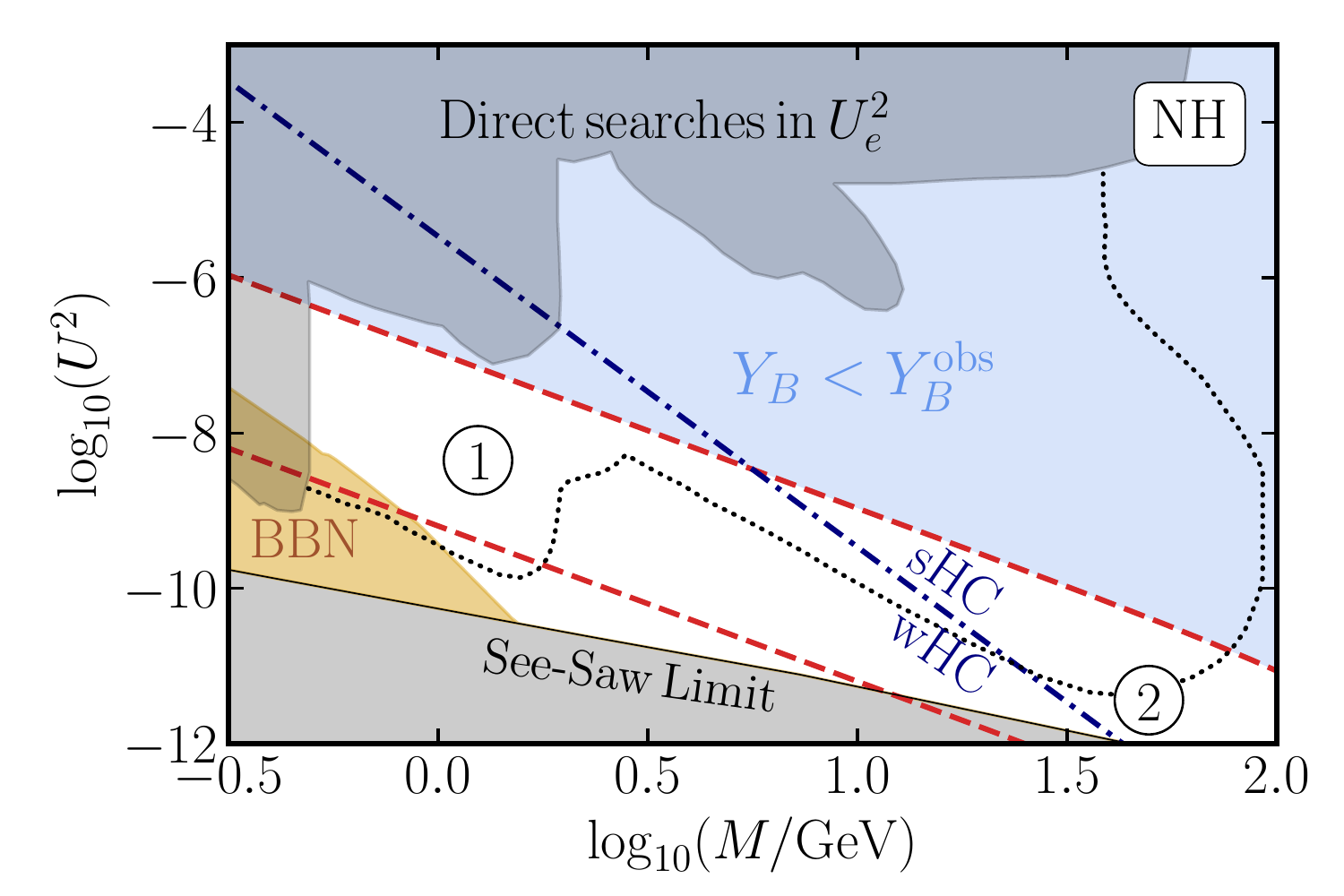} &
\hspace{-0.55cm}  \includegraphics[width=0.5\textwidth]{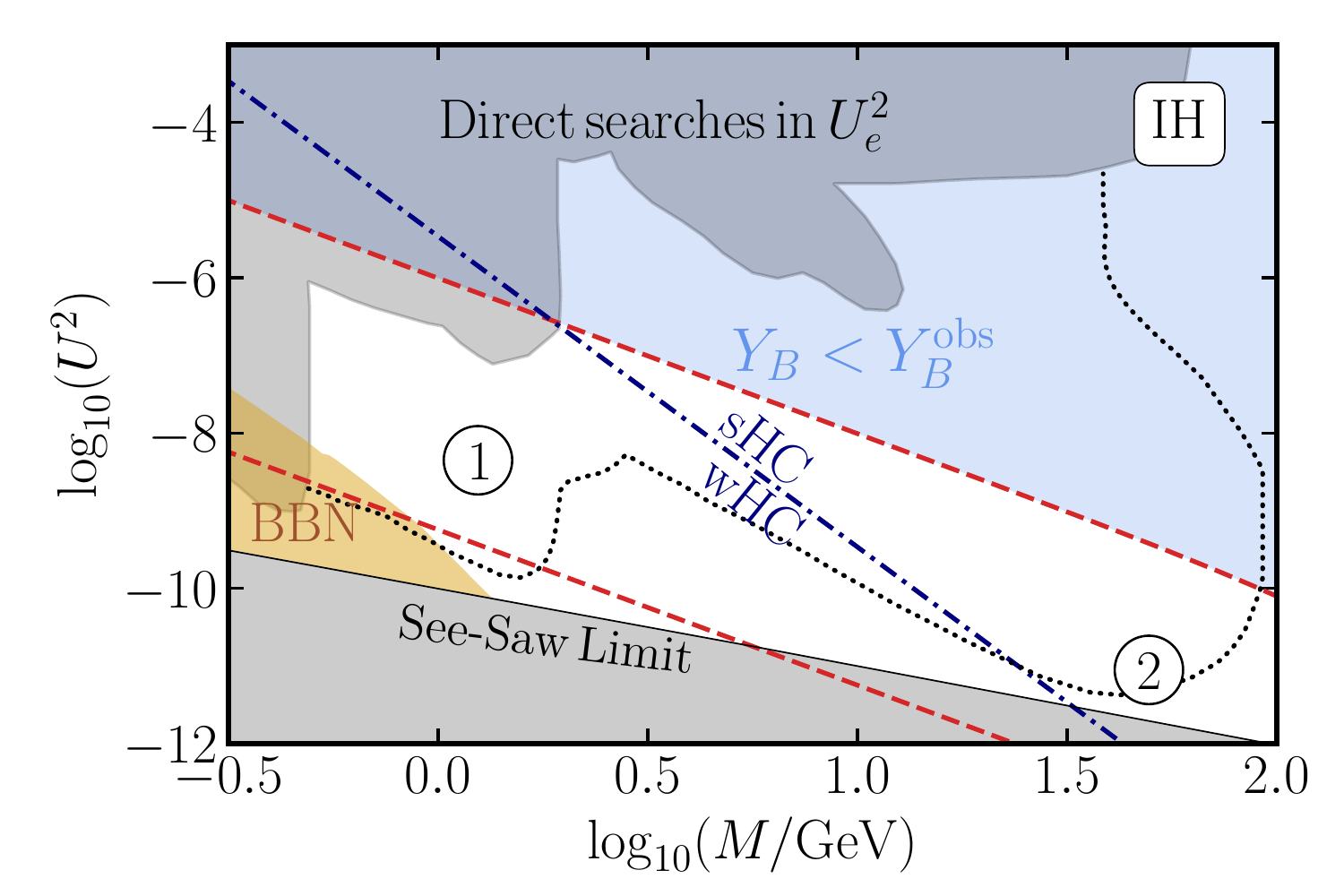}    
\end{tabular}
\vspace{-0.4cm}
\caption{
The two different weak washout regimes as discussed in the main text on the plane $(M, U^2)$ for NH (IH) in the left (right) panel.
Coloured regiones are excluded by either direct and indirect probes.
White regions fulfil all three Sakharov conditions and a baryon asymmetry can survive until today.
For reference, the region enclosed by the black dotted line marks the sensitivity region of SHiP~\cite{SHiP:2018xqw, Aberle:2839677}, MATHUSLA~\cite{MATHUSLA:2020uve} and FCC-ee~\cite{Blondel:2014bra, FCC:2018byv, FCC:2018evy}.
}
\label{fig:regimes}
\end{figure}
%
%
%
\begin{itemize}

\item Regime $1$ -- \textbf{Flavoured with wHC}.

\be
\label{eq:define_regime_4}
\Gamma_{\rm LN}(T_{\rm EW}), \Gamma_{M}(T_{\rm EW}), \Gamma_\alpha(T_{\rm EW}) < H_u(T_{\rm EW}) < \Gamma(T_{\rm EW}).
\ee

\item Regime $2$ -- \textbf{Flavoured with sHC}.

\be
\label{eq:define_regime_3}
\Gamma_{\rm LN}(T_{\rm EW}), \Gamma_\alpha(T_{\rm EW}) < H_u(T_{\rm EW}) < \Gamma_{M}(T_{\rm EW}),\Gamma(T_{\rm EW}).
\ee

\end{itemize}
We will now discuss the flavour-basis independent CP invariants that we expect to be relevant in each of these regimes.


\subsection{CP–violating flavour invariants and baryogenesis}
\label{subsec:cpinv_basic}
In the limit of $\Delta M \to 0$ the CP invariants considered in~\cite{Hernandez:2022ivz} vanish exactly.
However, non-zero CP invariants at higher order in the Yukawa couplings exist~\cite{Yu:2021cco, Drewes:2022kap}.
In the basis in which the charged lepton Yukawas $Y_l$ are diagonal, a non-vanishing contribution at leading order is
\begin{align}
\label{eq:cpinv_basic}
{\tilde I}_0 \equiv  {\rm Im} \left({\rm Tr}
\left[ Y^\dagger Y M_R^{*} Y^T  Y^{*} M_R Y^\dagger Y_{l }Y_{l}^\dagger Y \right] \right) \equiv \sum_\alpha y_{l_\alpha}^2 \Delta_\alpha,
\end{align}
where 
\begin{align}
\label{eq:cpinv_basic}
\Delta_\alpha =  {\rm Im}
\left[ \left( Y Y^\dagger Y M_R^{*} Y^T Y^{*} M_R Y^\dagger \right)_{\alpha \alpha} \right]\,.
\end{align}
Since $\sum_\alpha \Delta_\alpha = 0$, additional flavour effects are necessary to get a baryon asymmetry at this order. The baryon asymmetry generated in the two washout regimes will be proportional to different combinations of $\Delta_\alpha$. 
\vspace{0.25 cm}
\newline
\noindent
Regime $1$ -- \textbf{Flavoured with wHC.}
\noindent
If we assume that there is one weakly coupled flavour, $\alpha$, and the others are strongly coupled $\beta\neq \alpha$, we expect a contribution to the asymmetry of the former proportional to $\Delta_\alpha$ and a contribution from the latter weighted by $\Gamma_\beta^{-1}$  (see \cite{Hernandez:2022ivz}), so the net CP asymmetry will be a combination of two contributions, $\Delta^{\rm fw}_\alpha$ and $\Delta^{\rm M}_\beta$:
\begin{align}
 \Delta_{\alpha}^{\rm fw} &=  \frac{1}{\rm{Tr}\left( Y^\dagger Y \right)^2} \Delta_\alpha\,,
\label{eq:cpinv_flavoured_weak}
\end{align}
\begin{align}
 \sum_{\beta\neq \alpha} \Delta_\beta^{\rm M} &= \sum_{\beta\neq \alpha}  \frac{1}{\rm{Tr}\left( Y^\dagger Y \right)^2} \frac{\Delta_\beta}{\left( Y Y^\dagger \right)_{\beta\beta}}  \,,
\label{eq:cpinv_flavoured_strong}
\end{align}
where the matching to the analytical solution fixes the normalization factors, as we will see in Sec.~\ref{subsec:ana_sol}.
\vspace{0.25 cm}
\newline
\noindent
Regime $2$ -- \textbf{Flavoured with sHC.}
\noindent
In this flavoured case we expect the asymmetry to receive contributions only from the slow flavour, $\Delta_\alpha^{\rm fw}$, in Eq.~(\ref{eq:cpinv_flavoured_weak}). 
\vspace{0.25 cm}
\newline
\noindent
%
%
%
%


\section{CP invariants versus neutrino masses}
\label{sec:cpinv}

The CP invariants derived in the previous section can be expressed at leading order in $y'$ in terms of the parametrization of Eq.~\eqref{eq:LNVparam2N} as
\begin{align}
\label{eq:cpinv_flavoured_param}
\Delta_{\alpha}^{\rm fw} &=  \frac{2 M^2}{y^2} \sum_{\sigma \neq \alpha} y_\alpha y_\sigma y'_\alpha y'_\sigma \sin(\Delta\varphi_\sigma - \Delta\varphi_\alpha)\,,\\
\label{eq:cpinv_unflavoured_weak_param}
\Delta_\beta^{\rm M} &= \frac{2 M^2 }{y^2} \frac{y_\beta'}{y_\beta} \sum_{\sigma \neq \beta} y_\sigma y_\sigma' \sin\left( \Delta \varphi_\sigma - \Delta \varphi_\beta \right) \,,
\end{align}
with $\Delta \varphi = \varphi' - \varphi$.
On the other hand, the CP invariants can be related to the physical neutrino masses and other observable HNL parameters using the light neutrino mass constraint 
\be
- \left( m_{\nu}\right)_{\alpha\beta} = \frac{v^2}{\Lambda}\left( Y_{\alpha 1} Y_{\beta 2} + Y_{\alpha 2} Y_{\beta 1} \right)=\left(U_\nu^*m\,U_\nu^\dagger\right)_{\alpha\beta}\,.
\label{eq:mnuU}
\ee 
The parameters of the right handed neutrino Majorana mass matrix are related to the physical HNL masses simply by\footnote{The HNL masses get small corrections after the EWPT from the light neutrino masses that are $\mathcal{O}(m_\nu) \sim 10^{-11}\,\mathrm{GeV}$, see Eqs.~(\ref{eq:dMcorrectionsNH}) and~(\ref{eq:dMcorrectionsIH}).}
\bea
\Lambda \equiv M.\nonumber
\eea
The HNL flavour mixings are given by
\be
\label{eq:Hlmixing}
\Theta^*= YvM_R^{-1}W^*,
\ee
where $W$ is the unitary matrix which diagonalizes $M_R$, given by
\be
\label{eq:W}
W=\frac{1}{\sqrt{2}}\begin{pmatrix}
1 & 1\\
-1  & 1
\end{pmatrix}
{\rm diag}(i, 1)\,.
\ee
This leads to\footnote{Future colliders will not be able to distinguish the mixings of the two right handed neutrinos present in this model, but rather are sensitive to their average mixing. For this reason we change the definition of the mixing with respect to~\cite{Hernandez:2022ivz}.}
\begin{align}
\label{eq:U2}
U^2\equiv \frac{1}{2} \sum_{\alpha,I} |\Theta_{\alpha I}|^2 = \frac{y^2v^2}{2 M^2} \left[1 +\mathcal{O}\left(\frac{y'^2}{y^2}\right)\right]\,.
\end{align}
Note that all CP violation of the model arises from the Dirac ($\delta$) and Majorana ($\phi$) phases encoded in the PMNS matrix.

We will use as free parameters $(M, U^2, \delta,\phi)$. The Yukawa couplings can be written as a function of these and the 
 light neutrino mass differences and mixings~\cite{Gavela:2009cd}. 
The expressions differ for normal and inverted neutrino hierarchy, and are the same as reported in~\cite{Hernandez:2022ivz} when taking the limit of $\Delta M \to 0$.
For completeness we include the expressions in App.~\ref{app:param}.

These relations allow us to express the CP invariants of Eqs.~\eqref{eq:cpinv_flavoured_weak}-\eqref{eq:cpinv_flavoured_strong} in terms of observable quantities.
We summarize the results at leading order in $y'/y$ and expanding to leading order in the small light neutrino parameters
\begin{align}
\label{eq:CP_expansion}
r \equiv \frac{\sqrt{\Delta m_{\rm sol}^2}}{\sqrt{\Delta m_{\rm atm}^2}} \sim \theta_{13} \sim |\theta_{23} - \pi/4| \sim 10^{-1}\,.
\end{align}
\vspace{0.25 cm}
\newline
\noindent
\textbf{Normal hierarchy (NH):}\\
\begin{align}
\label{eq:CP_inv_U2_M_e_NH}
\Delta_{e}^{\rm fw} &= -\frac{ M^2 \Delta m_{\rm{atm}}^2 \sqrt{r}   }{2 U^2 v^2}~\theta_{13}  s_{12} \sin (\delta +\phi ) \,,\\
\label{eq:CP_inv_U2_M_mu_tau_NH}
\Delta_{\mu}^{\rm fw} &= -\Delta_{\tau}^{\rm fw} = - \frac{M^2 \Delta m_{\rm{atm}}^2  \sqrt{r}}{4 U^2 v^2}  c_{12} \sin\phi\,,\\
\label{eq:CP_inv_U2_M_M_e_NH}
\sum_{\beta \neq e} \Delta^{\rm M}_{\beta} &= \frac{ r\Delta m_{\rm{atm}}^2 }{2 U^4} ~ c_{12}^2 \sin(2\phi)\,,\\
\sum_{\beta \neq \mu} \Delta^{\rm M}_{\beta}  &=  \frac{\Delta m_{\rm{atm}}^2}{4 U^4\sqrt{r}}  ~ \frac{r c_{12} s_{12} \sin\phi - \theta_{13} \sin(\delta + \phi)  }{ s_{12}}   \, ,\\
\label{eq:CP_inv_U2_M_M_tau_NH}
\sum_{\beta \neq \tau} \Delta^{\rm M}_{\beta}  &=  -  \frac{\Delta m_{\rm{atm}}^2}{4 U^4\sqrt{r}}~ \frac{ r c_{12} s_{12} \sin\phi  + \theta_{13} \sin(\delta + \phi )  }{s_{12}}   \, .
\end{align}
\vspace{0.25 cm}
\newline
\noindent
\textbf{Inverted hierarchy (IH):}\\
\begin{align}
\label{eq:CP_inv_U2_M_e_IH}
\Delta_{e}^{\rm fw} &= \frac{ M^2 \Delta m_{\rm{atm}}^2 r^2  }{4 U^2 v^2} ~c_{12} s_{12} \sin\phi\,,\\
\label{eq:CP_inv_U2_M_mu_tau_IH}
\Delta_{\mu}^{\rm fw} &= \Delta_{\tau}^{\rm fw} = - \frac{1}{2} \Delta_{e}^{\rm fw} \,,\\
\label{eq:CP_inv_U2_M_M_e_IH}
\sum_{\beta \neq e} \Delta^{\rm M}_{\beta}  &=  \frac{\Delta m_{\rm{atm}}^2 }{2 U^4} \frac{r^2 c_{12} s_{12} \sin\phi}{2 c_{12} s_{12} \cos\phi - 1}  \,,\\
\label{eq:CP_inv_U2_M_M_mu_tau_IH}
\sum_{\beta \neq \mu} \Delta^{\rm M}_{\beta}  &= \sum_{\beta \neq \tau} \Delta^{\rm M}_{\beta} = \frac{\Delta m_{\rm{atm}}^2}{2 U^4} \frac{r^2 c_{12}^2 s_{12}^2 \sin(2\phi)}{4 c_{12}^2 s_{12}^2\cos^2(\phi) - 1} .
\end{align}

For some regions of parameter space, next-to-leading order corrections in the small parameters of Eq.~(\ref{eq:CP_expansion}) are non negligible, but they can be easily computed introducing Eqs.~(\ref{eq:Yno_PMNS})-(\ref{rhoIH}) in Eqs.~(\ref{eq:cpinv_flavoured_param}) and (\ref{eq:cpinv_unflavoured_weak_param}).


\section{Baryon asymmetry: kinetic equations and analytical approximations}
\label{sec:ana_approx}

\subsection{Kinetic equations}
The quantum kinetic equations that describe the generation of the baryon asymmetry have been studied in detail before (see for instance~\cite{Ghiglieri:2017gjz} for the complete derivation of the kinetic equations). 
We use the same equations as derived in~\cite{Hernandez:2016kel}, but adding the LNV corrections to the rates that have been computed in~\cite{Ghiglieri:2017gjz}. 
We have checked that they are equivalent to those in~\cite{Ghiglieri:2017gjz}, but neglecting the hypercharge chemical potential, which is a small effect. 
We consider only the momentum-averaged approximation, which reproduces the full momentum computation up to $\mathcal{O}(1)$ effects in the baryon asymmetry~\cite{Ghiglieri:2018wbs, Asaka:2011wq}.

We work in the basis in which the HNL mass matrix $M$ is diagonal.
We define the normalized heavy neutrino density matrices for the two helicities:
\begin{eqnarray}
r_N = {\rho_N \over \rho_F}, \;\;\;\; r_{\bar N} = {\rho_{\bar N}\over \rho_F}\,,
\end{eqnarray}
where  $\rho_F(z) =( \exp z + 1)^{-1}$ with $z=k/T$ is the Fermi-Dirac distribution.
The evolution of these matrices as a function of the scale factor $x  = a = T^{-1}$ is dictated by the equations:
\begin{eqnarray}
x H_u {\text{d} r_N\over \text{d} x} &=& -i [\langle H\rangle, r_N]  -{\langle\gamma^{(0)}_N\rangle\over 2} \{Y^\dagger Y, r_N-1\}-  x^2{\langle s^{(0)}_N\rangle\over 2} \{M Y^T Y^* M, r_N-1\}\nonumber\\
&+&  \langle \gamma_N^{(1)} \rangle Y^\dagger \mu Y  -  x^2 \langle s_N^{(1)} \rangle M Y^T \mu Y^* M \nonumber\\
&-& {\langle \gamma_N^{(2)}\rangle \over 2}  \big\{Y^\dagger \mu Y,r_N\big\} + x^2 \frac{\langle s_{N}^{(2)} \rangle}{2} \{M Y^T \mu Y^* M, r_N\} \,,\nonumber\\ 
x H_u {\text{d} r_{\bar N}\over \text{d} x} &=& -i [\langle H^*\rangle, r_{\bar N}]  -{\langle\gamma^{(0)}_N\rangle\over 2} \{Y^T Y^*, r_{\bar N}-1\} -x^2 {\langle s^{(0)}_N\rangle\over 2} \{M Y^\dagger Y M, r_{\bar N}-1\} 
\nonumber\\
&-&  \langle \gamma_N^{(1)} \rangle Y^T \mu Y^*   + x^2 \langle s_N^{(1)} \rangle M Y^\dagger \mu Y M \nonumber\\   
&+&  {\langle \gamma_N^{(2)}\rangle \over 2}  \big\{Y^T \mu Y^*,r_{\bar N}\big\} - x^2 \frac{\langle s_{N}^{(2)} \rangle}{2} \{M Y^{\dagger} \mu Y M, r_{\bar{N}}\} \,,\nonumber\\  
x H_u {\text{d} {\mu}_{B/3-L_\alpha}\over \text{d} x} & = & {\int_k \rho_F\over \int_k \rho'_F}  \left[{\langle \gamma_N^{(0)}\rangle \over 2} (Y r_N Y^\dagger- Y^* r_{\bar N} Y^T) - x^2 {\langle s_N^{(0)}\rangle \over 2} (Y^* M r_N M Y^T- Y M r_{\bar N} M Y^\dagger) \right.\nonumber\\
&-&\left.\mu_\alpha \left(\langle\gamma_N^{(1)}\rangle YY^\dagger +x^2 \langle s_N^{(1)}\rangle Y M^2Y^\dagger   \right)
 +   {\langle\gamma_N^{(2)}\rangle\over 2} \mu_\alpha (Y r_N Y^\dagger+Y^* r_{\bar N} Y^T)    \right. \nonumber\\
&+&\left. x^2 \frac{\langle s_{N}^{(2)}\rangle}{2} \mu_{\alpha}\left(YMr_{\bar{N}}MY^{\dagger} + Y^*Mr_{N}MY^{T} \right) \right]_{\alpha\alpha}\,,
\label{eq:rhonrhonbarav}
\end{eqnarray}
where $H_u(T)$ is the Hubble parameter of Eq.~(\ref{eq:hubble}) and $\rho_{F}'=\text{d}\rho_{F}/\text{d}z$. In these equations, the matrix $\mu\equiv {\mathrm diag}(\mu_\alpha)$ and $\mu_\alpha$  is the lepton chemical potential in flavour $\alpha$. $\mu_{B/3-L_\alpha}$ is related to the approximately conserved charge densities as:
\bea 
\label{eq:n_to_mu}
n_{B/3-L_\alpha} \equiv -2 \mu_{B/3 -L_\alpha} \int_k \rho'_F = {1\over 6} \mu_{B/3 -L_\alpha} T^3\,.
\eea
The relation between the two is 
\bea
\mu_{\alpha} &=& - \sum_\beta C_{\alpha\beta} \mu_{B/3-L_\beta}\,,
\eea
where the matrix $C$ is given by~\cite{Abada:2018oly}
\bea
C = -\frac{1}{711}\left(
\begin{array}{ccc}
257 & 20 & 20  \\
 20 & 257 & 20 \\
 20 & 20 & 257 \\
\end{array}
\right)\,.
\label{eq:Cmatrix}
\eea
The Hamiltonian term is given by
\begin{align}
&H(k_0, T) \equiv {M^2 \over 2 k_0} + V_N + V_N^M \,, \;\;\; V_N \equiv {T^2 \over 8 k_0} Y^\dagger Y\,, \\
\label{eq:H_VM}
&V_N^M \equiv \frac{Y^T Y^* T k_0}{ 16 k_0  \sqrt{k_0^2+M^2}} \left( \left(\sqrt{k_0^2+M^2}-k_0\right)- \frac{M^2}{2 k_0} \log \left(\frac{\sqrt{k_0^2+M^2}+k_0}{\sqrt{k_0^2+M^2}-k_0}\right) \right)\,. 
\end{align}
The LNC rates including  $1\leftrightarrow 2$ and $2\leftrightarrow 2$ processes have been expanded  to linear order in the leptonic chemical potential:
\begin{eqnarray}
\gamma_N(k,\mu_\alpha) \simeq \gamma_N^{(0)}+ \gamma_N^{(2)} \mu_\alpha\,,
\end{eqnarray}
while 
\bea
\label{eq:gamma1_definition}
\gamma_N^{(1)} \equiv \gamma_N^{(2)} - {\rho'_F\over \rho_F} \gamma_N^{(0)}\,.
\eea
The $s_N$ rates are expanded analogously. 
All the rates are momentum averaged:
\begin{eqnarray}
\label{eq:average_definition}
\langle (...)\rangle \equiv {\int_z (...) \rho_F(z)\over \int_z \rho_F(z)}\,.
\end{eqnarray}
The momentum dependent rates haven been derived to great level of detail in~\cite{Ghiglieri:2017gjz}.
We will use these rates for our numerical study.
Approximate expressions for the momentum averaged rates are given in~\cite{Hernandez:2022ivz} and summarized in Tab.~\ref{tab:rates}.
\begin{table}[!t]
\begin{center}
\begin{tabular}{|c|c|c|}
\hline
$n$ & $\langle \gamma^{(n)}_N(T)\rangle/T $ & $\langle s^{(n)}_N(T)\rangle/T$ \\
\hline
\hline
0 &  0.0091 & 0.0434  \\
1 &  0.0051 & 0.0086 \\
2 & -0.0022 & -0.0165\\
\hline
\end{tabular}
\caption{Coefficients in the momentum averaged rates at $T= 10^6$ GeV.}
\label{tab:rates}
\end{center}
\end{table}
Lastly, we define the factor
\begin{eqnarray}
{\int_k \rho_F\over \int_k \rho'_F} = -{9 \xi(3)\over \pi^2} \equiv -\kappa\,.
\end{eqnarray}
The initial conditions for the system is set by vanishing densities and chemical potentials at high temperatures, $x_{\mathrm{ini}} \sim 0 $.\footnote{Non-zero initial conditions have been studied e.g. in~\cite{Asaka:2017rdj}}
The system is then evolved from this initial state until the electroweak phase transition $x_{\mathrm{EW}}$.


\subsection{Perturbative Approximation }
\label{subsec:pert_approach}

In order to obtain a perturbative approximation to the solution of the Eqs.~\eqref{eq:rhonrhonbarav} we make the following approximations: i) neglect non-linear terms, ii) $x$-independent rates $(\gamma_i, s_i)$ and iii) diagonal $C$ matrix.

In the linearized limit, the quantum kinetic equation~\eqref{eq:rhonrhonbarav} can be written in the vector form
\begin{align}
\label{eq:vector_ode}
\frac{\mathrm{d} r(x)}{ \mathrm{d} x} = A(x) r(x) + h(x)\,,
\end{align} 
where we define the $11-$dimensional vector 
\bea 
r(x) &\equiv& \left([r_N]_{11}, [r_N]_{22}, {\mathrm{Re}}([r_N]_{12}), {\mathrm{Im}}([r_N]_{12}),[r_{\bar N}]_{11}, [r_{\bar N}]_{22}, {\mathrm{Re}}([r_{\bar N}]_{12}), {\mathrm{Im}}([r_{\bar N}]_{12}),\right.\nonumber\\
& & \left. \mu_{B/3-L_e},\mu_{B/3-L_\mu},\mu_{B/3-L_\tau}\right)\,.
\eea
The matrix $A(x)$ has dimension $11\times11$ and the source vector $h(x)$ has dimension $11 \times 1$.
Note that both quantities in general depend on $x$.
To find a general perturbative solution of Eq.~\eqref{eq:vector_ode} we proceed by expanding around the small parameters $y'$ and $M^2$
\begin{align}
\label{eq:A_expansion}
A(x) &= A_0 + A_1(x) + \mathcal{O}(y'^2, (M^2)^2)\,,\\
\label{eq:h_expansion}
h(x) &= h_0 + h_1(x) + \mathcal{O}(y'^2, (M^2)^2)\,,
\end{align}
where $A_n, h_n = {\mathcal O}(y'^n, (M^2)^n)$.
It is easy to show that at leading order $A$ and $h$ are $x-$independent.
The matrix $A_0$ can be diagonalized 
\be
A_0	= V_0 \lambda_0 V_0^{-1}\,,
\ee
in which we denote the eigenvector matrix of $A_0$ as $V_0$ and with $\lambda_0$ the diagonal eigenvalue matrix.
We note that contrary to the problem considered in~\cite{Hernandez:2022ivz} the eigenvectors are $x-$independent.
Therefore, the adiabatic approximation developed in~\cite{Hernandez:2022ivz} is exact and reduces to the standard perturbative approach.
It is then convenient to define the left and right matrix fundamentals
\begin{align}
\label{eq:matrix_fundamental}
\phi_{\mathrm{\ell}} = V_0 \mathrm{e}^{ \Lambda_0}\,,\,\,\,\, \phi_{\mathrm{r}} = {e}^{ - \Lambda_0} V_0^{-1}\,,
\end{align}
with
\be
\label{eq:int_eigenvalue}
\Lambda_0(x) \equiv \int^x_0 \mathrm{d}z\, \lambda_0(z)\,.
\ee
At leading order the result can be expressed as
\begin{align}
\label{eq:sol_q0}
r^{(0)} = \phi_{\mathrm{\ell}}(x) \int_0^{x} \mathrm{d}z\, \phi_{\mathrm{r}}(z) h_0(z) \equiv \phi_{\mathrm{\ell}}(x) I_0(x)\,.
\end{align}
The appearing $1-$dimensional integral can be evaluated analytically for $n \in \mathbb{N} $ and yields
\begin{align}
\label{eq:integral_evaluation}
\int_0^x \mathrm{d}t \, t^n {e}^{ \alpha + \beta t} = \frac{n! {e}^{\alpha}}{\beta^{n+1}} \left( 1 + {e}^{\beta x} \sum_{i=0}^{n}  \frac{ \left( -1 \right)^{n + i}}{i!} \beta^i x^i \right)\,.
\end{align}
At first order in the perturbations of $y'$ and $M^2$ the correction satisfies the equation
\begin{align}
{\mathrm{d} r^{(1)}\over \mathrm{d} x} = A_0 r^{(1)} + A_1 r^{(0)} + h_1\,.
\end{align}
Its solution is found by
\begin{align}
\label{eq:r1}
r^{(1)}= \phi_{\mathrm{\ell}}(x) \int_0^x {d}z \left[ \phi_{\mathrm{r}}(z) A_1(z) \phi_{\mathrm{\ell}}(z)I_0(z) + \phi_{\mathrm{r}}(z) h_1(z) \right] \,.
\end{align}
To evaluate the integral of Eq.~\eqref{eq:r1} we can, again, use Eq.~\eqref{eq:integral_evaluation}.
Higher order corrections are obtained in the same manner.
The first non-zero contribution to the asymmetry arises at third order and is of $\mathcal{O}(y'^2 M^2)$, as expected from the CP invariant discussion.


\subsection{Thermalization rates}
\label{subsec:thermalization_rates}

Let us now look at the thermalization rates. For the quantum kinetic equation of the system as expressed in Eq.~\eqref{eq:vector_ode}, the thermalization rates are given by the real parts of the time integrated eigenvalues of $A(x)$. 
It is actually easy to show that for $x\to\infty$ the system reaches statistical thermal equilibrium, i.e.
\be
\lim_{x\rightarrow \infty} r(x) = (1,1,0,0,1,1,0,0,0,0,0)\,,
\ee
because all eigenvalues of $A$ have negative real parts.
As naively expected by Eq.~\eqref{eq:matrix_fundamental} and Eq.~\eqref{eq:sol_q0} this limit is approached exponentially fast, i.e.
\begin{eqnarray}
\propto e^{- \Lambda_i(x)} \equiv \exp\left({-\int_0^x \mathrm{d}z  |{\mathrm{Re}}(\lambda_i(z))|}\right)\,,
\end{eqnarray}
 with $\lambda_i$ the eigenvalues of $A(x)$. 
To explicitly evaluate this expression it is convenient to normalize $x$ such that at $T= T_{\mathrm{EW}}$ we have $x_{\mathrm{EW}} = 1$ and introduce the variables
\begin{align}
 \label{eq:defs}
\gamma_i \equiv {\langle \gamma^{(i)}\rangle\over T}  {M_P^*\over T_{\mathrm{EW}}},  ~s_i \equiv {\langle s^{(i)}\rangle\over T}  {M_P^*\over T_{\mathrm{EW}}}, ~\omega\equiv {c_H\over 8} {M_P^*\over T_{\mathrm{EW}}},~ \omega_M \equiv \frac{M_{\mathrm{P}}^{*}}{T_{\mathrm{EW}}} \langle V_N^M\rangle x^{-2} \frac{T_{\mathrm{EW}}^2}{M^2}\,,
\end{align}
with
\bea
c_H\equiv  {\pi^2 \over 18 \zeta(3)}\,.
\eea
The momentum average of the mass dependent thermal mass contribution to the Hamiltonian, see Eq.~\eqref{eq:H_VM}, can be approximated as~\cite{Antusch:2017pkq}
\begin{align}
\label{eq:V_N_M}
\langle V_N^M \rangle \simeq x^2 \frac{M^2}{T_{\mathrm{EW}}^2}2.5\times10^{-3} \left(3.5 - 0.47 \log\left( \frac{M^2}{T_{\mathrm{EW}}^2} x^2 \right) + 3.46 \log\left( \frac{M}{T_{\mathrm{EW}}} x \right) \right)\,.
\end{align}
Therefore, $\omega_M$ is at leading order $x-$ and $M-$independent and is approximately
\be
\omega_M \simeq {M_P^*\over T_{\mathrm{EW}}} 8.75\times10^{-3}\,.
\ee
The largest real part corresponds to the fastest thermalization rate, that we can identify with $\Gamma$:
\begin{eqnarray}
\label{eq:Lambdamax}
\Lambda^{\mathrm{max}}(x) =\int_0^x \text{d} z~ {\mathrm{Max}}(|{\mathrm{Re}}(\lambda(z))|)  = y^2 \gamma_0 x \equiv \int_0^x \text{d}z ~{\Gamma\over z H_u}\,.
\end{eqnarray}
In the testable parameter space, $\Lambda^{\mathrm{max}}(x_{\mathrm{EW}}) > 1$.
Of more interest, however, are the thermalization rates associated to the regimes discussed in Sec.~\ref{subsec:sakharov}.
These are in general associated to the eigenvalues with the smallest real part.
Full thermalization is prevented when one the slow modes satisfies
\bea
\Lambda_i(x_{\mathrm{EW}}) \leq 1\,.
\label{eq:noneq}
\eea
We discuss in the following the three slow modes of the system $i=(\alpha, M, \mathrm{LN})$.

In the flavoured weak washout region, a slow mode remains in flavour $\alpha$ provided there is a hierarchy in the yukawas $\epsilon_\alpha = y_\alpha/y \ll 1$. 
The slow rate of the flavoured weak washout regime is identified from the  corresponding eigenvalue
\begin{eqnarray}
\label{eq:Gamma_alpha}
\Lambda_\alpha(x) = \frac{\left( 6 \gamma_0 s_0 + \gamma_1 s_0 \kappa + \gamma_0 s_1 \kappa \right) \left( 3 \gamma_1 x + (M/T_{\mathrm{EW}})^2 s_1 x^3 \right)}{6 \left( 4 \gamma_0 s_0 + \gamma_1 s_0 \kappa + \gamma_0 s_1 \kappa \right)} \kappa y_\alpha^2 \equiv \int_0^x \text{d} z {\Gamma_\alpha(z)\over z H_u(z)}\,.\,\,\,\,\,
\end{eqnarray}
The boundary of the weak flavour washout region is therefore
\begin{eqnarray}
\Lambda_{\alpha}(x_{\mathrm{EW}}) = 1\,,
\label{eq:wLNV}
\end{eqnarray}
The condition for the less coupled flavour defines the upper limit of the dashed band in Fig.~\ref{fig:regimes}.
Note that for small helicity conserving rates Eq.~\eqref{eq:Gamma_alpha} can be approximated by 
\be
\Lambda_\alpha(x)  \simeq {3\over 4} y_\alpha^2 \kappa \gamma_1 x\,,
\ee
which is a useful estimate for most of the range of masses considered here. 

The second slow mode is related to the helicity conserving $\propto M/T$  interaction rates.
The corresponding eigenvalue is
\begin{eqnarray}
\label{eq:Gamma_slow_M_ov}
\Lambda_{\mathrm{M}}(x) =  {1\over 3} {M^2\over T_{\mathrm{EW}}^2} x^3 s_0 y^2 \equiv \int_0^x \text{d} z {\Gamma ^{\mathrm{slow}}_M(z)\over z H_u} \,.
\end{eqnarray}
The boundary of the wHC/sHC regions (dashed-dotted line in Fig.~\ref{fig:regimes}) is defined by the condition
\begin{eqnarray}
\Lambda_M(x_{\mathrm{EW}}) = 1\,.
\end{eqnarray}

The last slow mode is associated to the breaking of the generalized LN with the assignment given in Eq.~\eqref{eq:LNtexture}.
Because the two HNLs are assumed to be exactly degenerate the slow mode is only non-zero when including $y'$.
Indeed one finds
\begin{eqnarray}
\label{eq:Gamma_slow_LN}
\Lambda_{\mathrm{LN}}(x) = y'^2 \gamma_0 x  \equiv \int_0^x \text{d} z {\Gamma ^{\mathrm{slow}}_{\mathrm{LN}}(z)\over z H_u} \,.
\end{eqnarray}
The boundary is again defined by
\begin{eqnarray}
\label{eq:LN_boundary}
\Lambda_{\mathrm{LN}}(x_{\mathrm{EW}}) = 1\,.
\end{eqnarray}


\subsection{Projection method for strong helicity conserving interactions}
\label{subsec:projection}

The perturbative approximation assumes an expansion in $M/T$, which means that it is valid only for times, $x$, such that 
$\Gamma_M(x) \leq H_u(x)$, or $\Lambda_M(x) \leq 1$.  This means that the perturbative approach is valid for $x\leq x_{\rm EW}$ within the wHC regimes, flavoured or unflavoured, but not in the sHC ones.

In the latter  case, a good approximation can be obtained as follows. We evolve the perturbative solution up to some $x_M$ such that $\Lambda_M(x_{\mathrm{M}}) = 1$. Using Eq.~\eqref{eq:Gamma_slow_M_ov} we find
\be
\label{eq:xM}
x_{\mathrm{M}}= \left( {3 T_{\mathrm{EW}}^2 \over  M^2 s_0  y^2}\right)^{1/3}\,.
\ee
For $x \geq x_M$ the asymmetry reaches a quasi stationary solution.
The final baryon asymmetry can therefore be obtained by projecting the solution vector $r_N(x_{\mathrm{M}})$ onto the approximate zero mode of the slow flavour $\alpha$.
This can be done as follows.
Let us denote by $v_i\,(w_i)$ the right (left) eigenvectors of $A$, which satisfy the orthonormality relation $w_i^\dagger v_j =\delta_{ij}$.
If at the times $ x \geq x_{\mathrm{M}}$ all modes are strongly coupled to the plasma except one mode\footnote{This will be in practice the slow flavour mode $\alpha$ if it exists. Even though the LN mode is always weakly coupled, the asymmetry in this mode is  \textit{exactly} zero at order $\mathcal{O}(y'^2)$ and therefore we neglect it here}, we can denote with $v_0\,(w_0)$ the right (left) eigenvectors of this weakly coupled mode.
Then, the solution at $x\geq x_M$ can be approximated  by
\begin{eqnarray}
\label{eq:onezm}
r(x) \simeq \left(w_0^\dagger\cdot r(x_{\rm M}) \right) v_0\,.
\end{eqnarray}
If there are more modes, the result is the sum of the projection on each zero mode.

More details on the derivation of Eq.~\eqref{eq:onezm} can be found in~\cite{Hernandez:2022ivz}.


\subsection{Analytical  results}
\label{subsec:ana_sol}

All analytical results are expressed in terms of the CP invariants as derived in Sec.~\ref{sec:cpinv}.
In terms of the parameters of Eq.~(\ref{eq:LNVparam2N}) they are given in Eqs.~(\ref{eq:cpinv_flavoured_param})-(\ref{eq:cpinv_unflavoured_weak_param}).
Their relation to physical observable quantities is given in Eqs.~(\ref{eq:CP_inv_U2_M_e_NH})-(\ref{eq:CP_inv_U2_M_M_tau_NH}) (Eqs.~(\ref{eq:CP_inv_U2_M_e_IH})-(\ref{eq:CP_inv_U2_M_M_mu_tau_IH})) for NH (IH).

\noindent
\subsubsection{Regime 1 -- flavoured wHC}
As long as $\Gamma_M(T_{\rm EW}) \leq H_u(T_{\rm EW})$, the perturbative solution is a good approximation at any $x \leq x_{\rm EW}$. This condition is satisfied in regime 1. The result for the late time $B-L$ chemical potentials at $x$ is
\begin{align}
\sum_{\alpha } \mu_{B/3 - L_\alpha} &= - \frac{x^2}{T_{\mathrm{EW}}^2} \frac{4 \gamma_0 \kappa (s_0 \omega + \gamma_0 \omega_M)}{(4 \gamma_0 + \gamma_1 \kappa)(\gamma_0^2 + 4 \omega^2)} \left( \frac{\gamma_1 \kappa x \Delta^{\mathrm{fw}}_{\alpha} }{3} + 2 \sum_{\beta \neq \alpha} \Delta^{\mathrm{M}}_\beta \right)\,,
\label{eq:ana_sol_cp_inv_fw_wlnv}
\end{align}
This result nicely matches the expectation of Eqs.~(\ref{eq:cpinv_flavoured_weak}) and (\ref{eq:cpinv_flavoured_strong}). If two flavour $(\alpha_1, \alpha_2)$ happen to be slow, we can use the same formula and just replace $(4 \gamma_0 + \gamma_1 \kappa) \mapsto (2 \gamma_0 + \gamma_1 \kappa)$ and sum over the slow flavours.

\noindent
\subsubsection{Regime 2 -- flavoured sHC}
\label{subsec:ana_approx_sHC}
Contrary to the previous regime we now have $\Gamma_M(T_{\rm EW}) > H_u(T_{\rm EW})$.
The perturbative solution is evolved until $x_M$, defined in Eq.~\eqref{eq:xM}, and then projected onto the slow mode(s). 
We find that the projection on the LN mode vanishes at ${\mathcal O}(y'^2)$ and is therefore negligible, but there is a contribution from the slow $\alpha$ mode. 
The result for the late time $B-L$ chemical potentials is
\begin{align}
\label{eq:asym_fw_slnv}
\sum_\alpha \mu_{B-L_{\alpha/3}} = - x_M^3 \frac{1}{T_{\mathrm{EW}}^2} \frac{8 \gamma_0 \kappa^2 (\gamma_1 s_0 + \gamma_0 s_1)(s_0 \omega + \gamma_0 \omega_M)}{6 (4\gamma_0 s_0 + \gamma_1 s_0 \kappa + \gamma_0 s_1 \kappa) (\gamma_0^2 + 4 \omega^2)} \Delta_{\alpha}^{\mathrm{fw}}\,,
\end{align}
For the scenario with two slow flavour $(\alpha_1,\alpha_2)$ we can use the same formula but replacing $\Delta_{\alpha}^{\mathrm{fw}} \mapsto 1/2 \sum_\alpha \Delta_{\alpha}^{\mathrm{fw}}$.

\subsubsection{Relating to the baryon asymmetry}
\label{subsec:mu_to_yb}
The analytical solutions for the $B-L$ chemical potentials derived above are directly related to the baryon asymmetry.
If one assumes the sphaleron rate to be suppressed instantaneously at $T=T_{\mathrm{EW}}$, it is a well known result that~\cite{Laine:1999wv, Harvey:1990}
\be
\label{eq:Yb_inst}
Y_B^{\mathrm{inst}} \simeq 1.26\times10^{-3} \sum_\alpha \mu_{B/3 - L_\alpha}\,.
\ee
This approximation is sufficiently good as long as one mode remains weak, which all our analytical solutions assume.
On the other hand, if all relevant modes enter the strong washout close to $x_{\mathrm{EW}}$, the final baryon asymmetry becomes sensitive to the exact dynamics of the EWPT.
In particular, employing a gradual freeze-out of the sphalerons, the prediction for the baryon asymmetry can differ up to $\mathcal{O}(10)$ compared to the naive estimate of Eq.~\eqref{eq:Yb_inst}~\cite{Hernandez:2022ivz, Eijima:2017cxr}.
This is mainly due to two effects: i) it restrains the impact of the significant growth of the helicity conserving rates within the broken Higgs phase on the baryon asymmetry and ii) it leads to a general reduction of the washout of the asymmetry.
For the numerical analysis we have implemented the smooth freeze-out of the sphalerons as outlined in~\cite{Eijima:2017cxr}.
Finally, we quote the experimentally measured value of the baryon asymmetry~\cite{Planck:2018vyg} that we have used
\be
Y_B^{\mathrm{exp}} = (8.66 \pm 0.05) \times 10^{-11}\,.
\ee

\subsection{Analytical results versus numerical solutions}
\label{subsec:ana_vs_num}

The analytical solutions above are leading order asymptotic solutions for $\sum_\alpha \mu_{B/3 - L_\alpha}$. To verify the accuracy of the analytical solutions we compare them to the numerical solutions for the differential equations obtained under two separate conditions: i) under the same approximations (i.e. linearized, constant rates and diagonal $C$), and ii) without any approximation. 
This is most easily done by using the parametrization of Eq.~\eqref{eq:LNVparam2N}, which does not include the constraints from neutrino oscillations data.
The inclusion of the latter just lead to non-linearly correlated Yukawa couplings (see Eq.~\eqref{eq:mnuU}), but does not change anything qualitatively.
In Tab.~\ref{tab:ana_vs_num_initial_condition} we summarize the chosen model parameters corresponding to the two regimes. 

\begin{figure}[!t]
\centering
\begin{tabular}{cc}
\hspace{-0.5cm}  \includegraphics[width=0.49\textwidth]{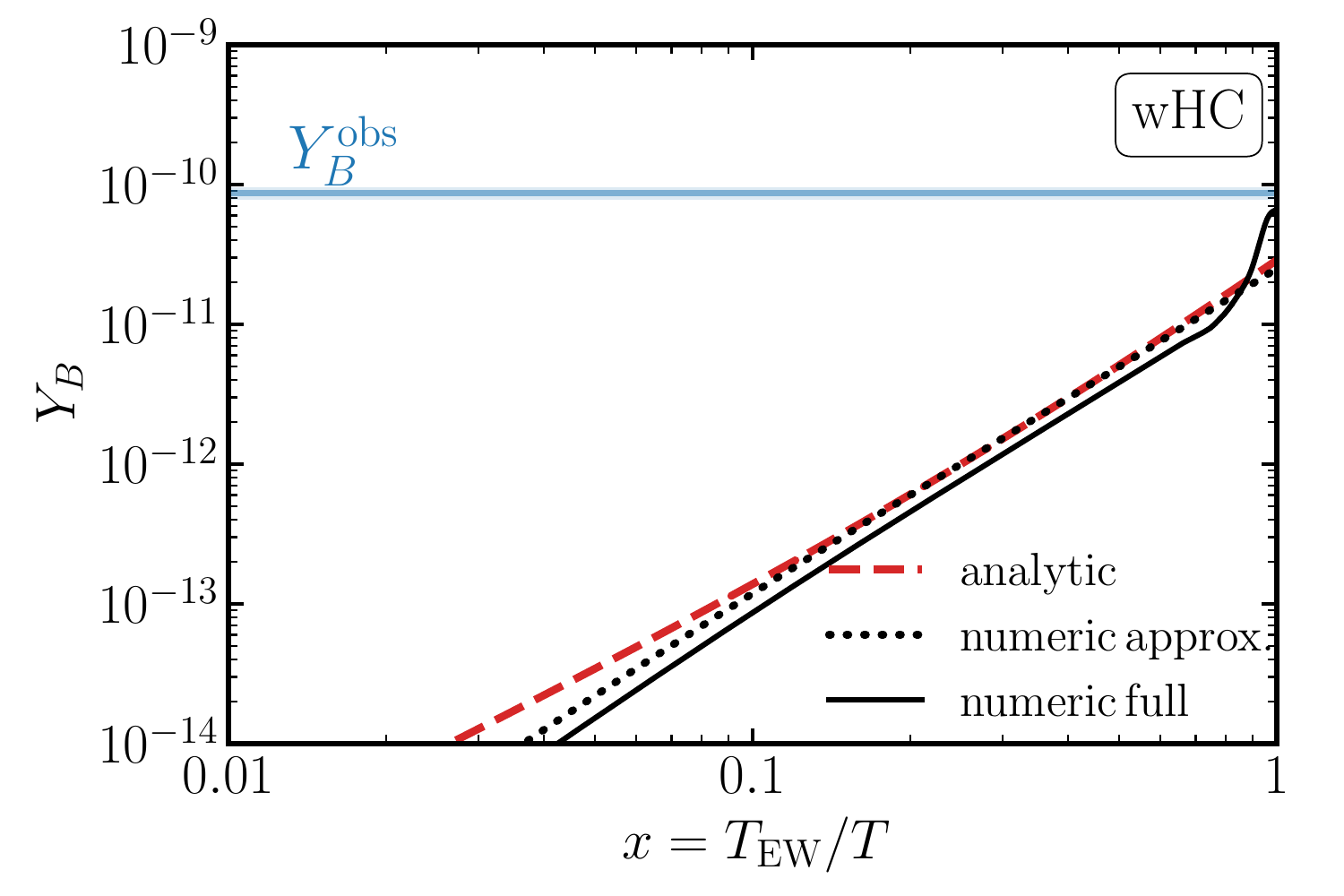} &
\hspace{-0.5cm}  \includegraphics[width=0.49\textwidth]{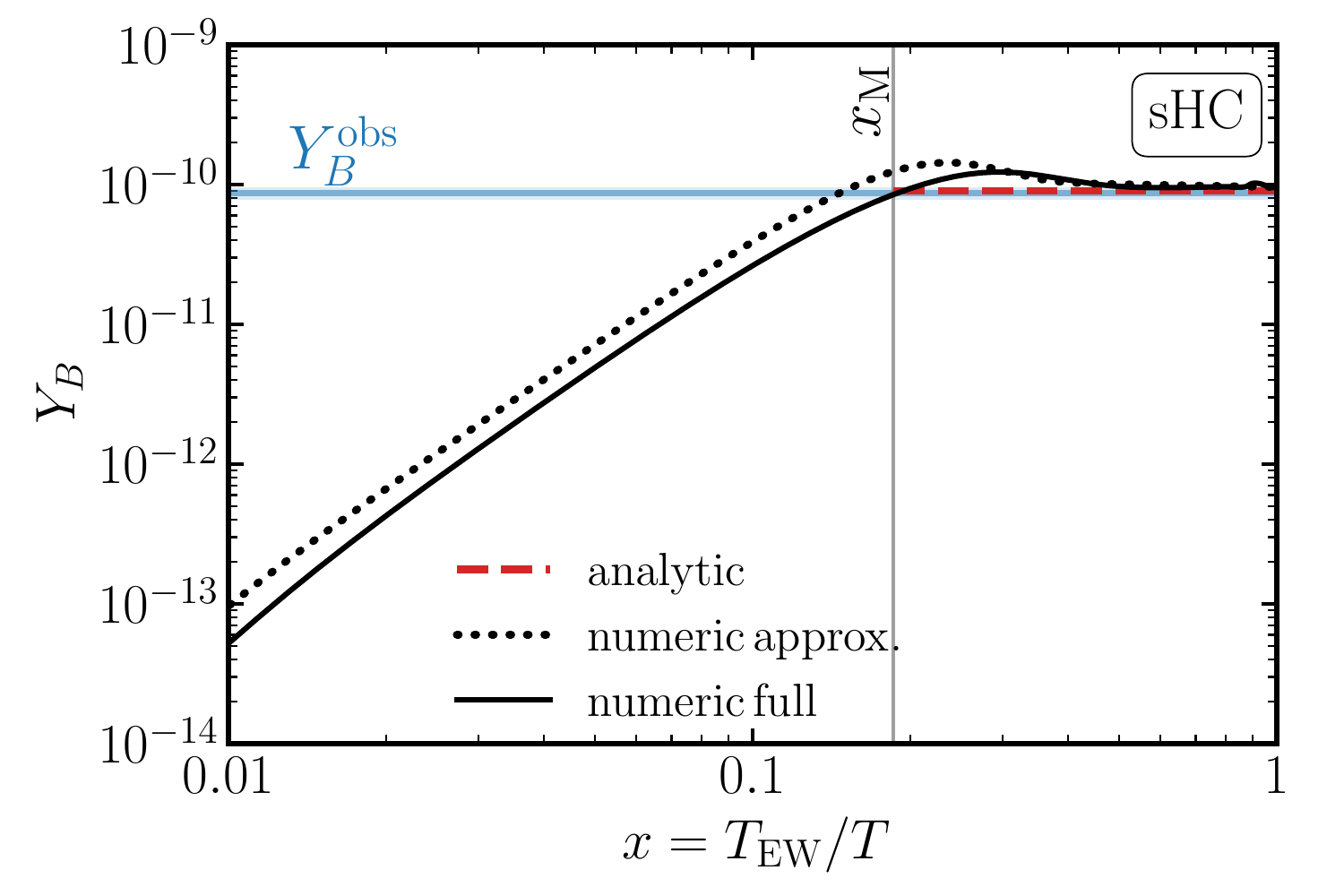} 
\end{tabular}
\vspace{-0.4cm}
\caption{ 
Comparison of analytical and numerical solutions in the wHC (left) and sHC (right) regimes for the parameters of Tab.~\ref{tab:ana_vs_num_initial_condition}. 
The corresponding analytical solutions of Eq.~\eqref{eq:ana_sol_cp_inv_fw_wlnv} and Eq.~\eqref{eq:asym_fw_slnv} are shown in dashed blue.
In solid (dotted) black we show the corresponding numerical solutions of the full (approximate) system, see main text.
The vertical grey line in the right panel indicates the projection time when $\Lambda^M = 1$ given by Eq.~\eqref{eq:xM}.
For reference, we indicate the observed value of the baryon asymmetry with the horizontal blue band.
}
\label{fig:ana_vs_num}
\end{figure}

In  Fig.~\ref{fig:ana_vs_num} we show the comparison of the analytical solution and the two numerical approximations. 
%
%
\begin{table}[!t]
\begin{center}
\begin{tabular}{| c | c c c c  c c c c |}
\hline
Regime & $M/\rm{GeV}$ & $\log_{10}(y_{e})$ & $\log_{10}(y_{\mu})$ & $\log_{10}(y_{\tau})$ & $\log_{10}(y'_{\alpha})$ &  $\Delta \varphi_e$ & $\Delta \varphi_\mu$ & $\Delta \varphi_\tau$ \\
\hline
\hline
1 (wHC) & 1 & $-7.2$ & $-6$  & $-5.7$ & $-7.5$ &  $\pi/2$ & $\pi/3$ & $\pi/4$  \\
2 (sHC) & 100 & $-7.2$ & $-6$  & $-5.8$ & $-8$ &  $\pi/2$ & $\pi/3$ & $\pi/4$  \\
\hline
\end{tabular}
\caption{
Model parameters for the comparison between the analytical and numerical solutions shown in Fig.~\ref{fig:ana_vs_num} for the regimes as labelled in Fig.~\ref{fig:regimes} and defined in Eqs.~\eqref{eq:define_regime_4}-\eqref{eq:define_regime_3}.
}
\label{tab:ana_vs_num_initial_condition}
\end{center}
\end{table}
%
%
%
%
%
%
%
We see that the analytical solution nicely reproduces the large-$x$ dependence of the numerical solution under the same approximations\footnote{The simple solutions we quote assume that all the modes that satisfied $\Lambda_i(x_{\rm EW}) > 1$, are fully washout. Of course this might not be the case at $x < x_{\rm EW}$, so we do not expect the results to match the numerical solution at early times. The perturbative solution does  provide accurate results in this regime but the expressions would be too long and not illuminating.}, 
while it reproduces the full numerical solution within a $\mathcal{O}(1)$ uncertainty in the wHC regime (left panel). 
The deviation is mainly due to the assumption of constant rates in the analytical approach. 
In particular, the full temperature dependent rates show a strong enhancement near  the EWPT, which, however, is partially absorbed by the smooth sphaleron freeze-out, for a detailed discussion see Ref.~\cite{Hernandez:2022ivz}. 
On the other hand, in the sHC regime (right panel) the asymmetry reaches a quasi-stationary state at temperatures above the EWPT and the final result is insensitive to the variation of the rates later on, so the approximation is more reliable. This is actually the regime that can be tested at FCC-ee (see Fig.~\ref{fig:regimes}).



\section{Parameter constraints from the baryon asymmetry}
\label{sec:parambounds}

The analytical results of the previous section predict the baryon asymmetry in terms of observable parameters.
In particular, the asymmetry can be expressed as a function of the HNL mass $M$, its mixing $U^2$ and the two CP-violating phases $(\delta, \phi)$.
It is therefore natural to ask what constraints successful baryogenesis imposes on these observables.
In this section we address this question.
All the findings will then be compared in the next section to the full numerical analysis.
For concreteness, we will use the instantaneous sphaleron freeze-out approximation of Eq.~\eqref{eq:Yb_inst} and evaluate the interaction rates at $T=10^6\,\rm{GeV}$ for $M=1\,\rm{GeV}$, see Tab.~\ref{tab:rates}.

\begin{figure}[!t]
\centering
\begin{tabular}{cc}
\hspace{-0.5cm}  \includegraphics[width=0.49\textwidth]{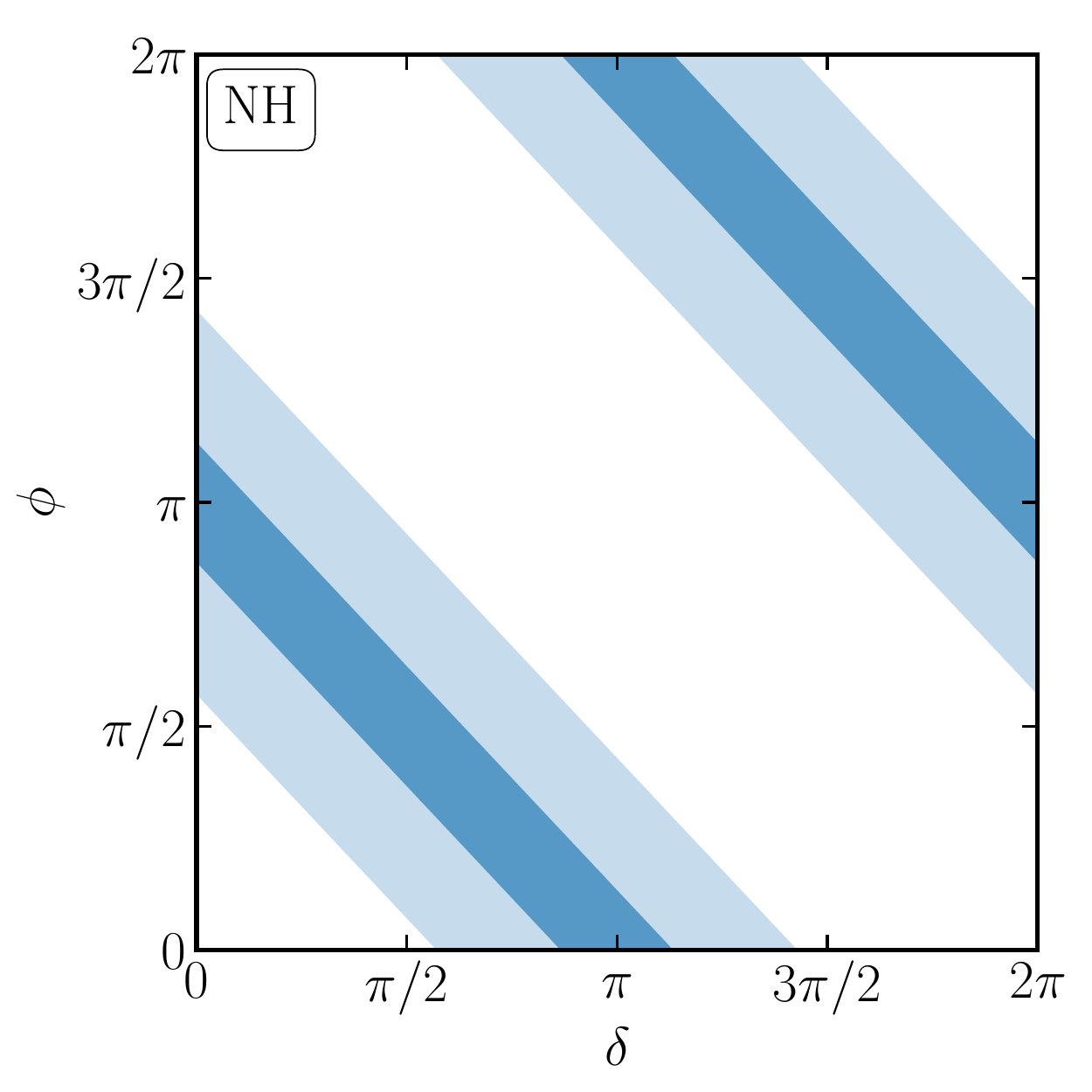} &
\hspace{-0.5cm}  \includegraphics[width=0.49\textwidth]{ 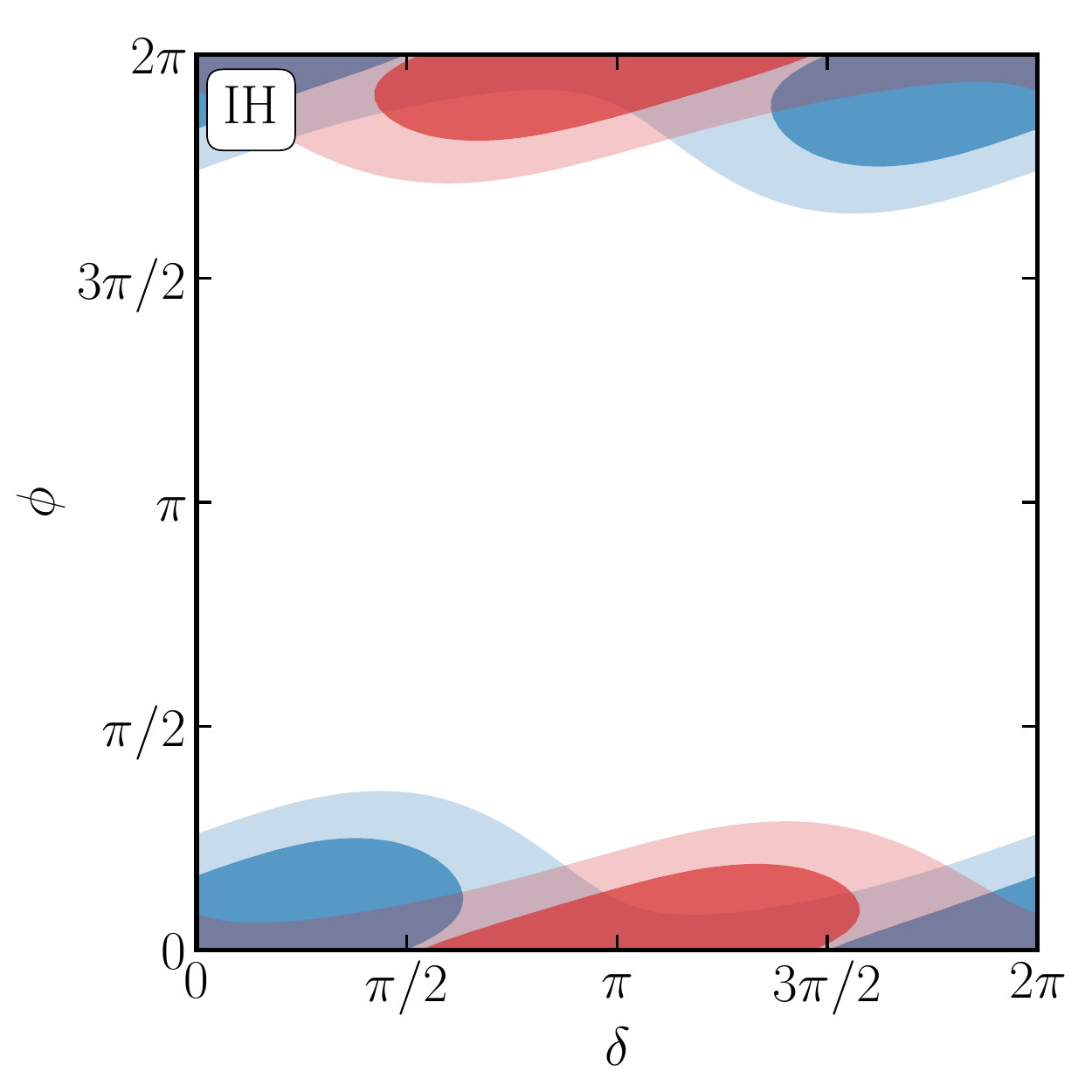} 
\end{tabular}
\vspace{-0.4cm}
\caption{ 
Contour regions of $\epsilon_\alpha$ for the relevant flavours in $\rm{NH}$ ($\rm{IH}$) in the \textit{left} (\textit{right}) panel.
\textit{Left}: The dark (light) blue regions correspond to $\epsilon_e \leq 0.01$ ($\epsilon_e \leq 0.05$).
\textit{Right}: Two flavours are relevant, the $\mu$ and the $\tau$. The dark (light) blue regions correspond to $\epsilon_\mu \leq 0.03$ ($\epsilon_\mu \leq 0.07$) and the dark (light) red regions correspond to $\epsilon_\tau \leq 0.03$ ($\epsilon_\tau \leq 0.07$).
}
\label{fig:f_eps_fw_ov}
\end{figure}

Successful baryogenesis in the regime $\Gamma(T_{\mathrm{EW}}) > H_u(T_{\mathrm{EW}})$ requires a flavour hierarchy in the Yukawas.\footnote{For the derivation showing that the asymmetry can not be explained in the unflavoured region, see App.~\ref{app:unflavoured}.}  
This means that (at least) one flavour has to remain out of equilibrium at $x_{\rm{EW}}$, while at the same time having at least one strongly coupled flavour.
The corresponding modes therefore need to fulfil $\Lambda_\alpha(x_{\rm{EW}}) \leq 1 \land \Lambda_\beta(x_{\rm{EW}}) \geq 1$, see Eq.~\eqref{eq:Gamma_alpha}, for some flavours $\alpha,\, \beta$.  
This leads to the \textit{necessary}, but not sufficient, condition
\be
\label{eq:U2_fw}
 10^{-9} \left({1{\rm GeV}\over M}\right)^2{1\over {\rm Max}(\epsilon_\alpha)} \lesssim U^2 \lesssim   10^{-9} \left({1{\rm GeV}\over M}\right)^2{1\over {\rm Min}(\epsilon_\alpha)}\,.
\ee
$\epsilon_\alpha = y_\alpha^2/y^2$ depends only on the PMNS parameters and in particular the unknown CP phases, $(\delta, \phi)$.  
Its minimum is given by
\begin{eqnarray}
\label{eq:minepsalpha}
{\rm Min}(\epsilon_e)_{NH} \simeq  {\rm Min}(\epsilon_\tau)_{IH} \simeq 10 \times {\rm Min}(\epsilon_\mu)_{IH} \simeq 5\times 10^{-3}  \,.
\end{eqnarray}
The maximum has only a mild dependence on the light neutrino mass hierarchy and is in both cases $\mathcal{O}(1)$.
This defines the region between the red dashed lines in Fig.~\ref{fig:regimes}.
In Fig.~\ref{fig:f_eps_fw_ov}, we show contour lines of small values of $\epsilon_\alpha$ for the relevant flavours both in NH and IH.
%
%
%
%
%
%
The generation of the baryon asymmetry inside the region defined in Eq.~\eqref{eq:U2_fw} depends on whether the helicity conserving rates are weak (regime 1) or strong (regime 2).

\subsection{Regime $1$ -- flavoured wHC}
\label{subsec:parambounds_wHC}
To prevent the helicity conserving interactions from equilibrating, $\Lambda_M(x_{\rm{EW}}) \leq 1$, see Eq.~\eqref{eq:Gamma_slow_M_ov}, the HNL mixing needs to obey
\be
U^2 \leq 1 \times 10^{-6} \left( \frac{1\,\rm{GeV}}{M} \right)^{4}\,.
\ee
This corresponds to the HNL mixing being below the blue dashed-dotted  line in the Fig.~\ref{fig:regimes}.
Using Eq.~\eqref{eq:ana_sol_cp_inv_fw_wlnv} the baryon asymmetry, together with the CP invariants from Eqs.~\eqref{eq:CP_inv_U2_M_e_NH}-\eqref{eq:CP_inv_U2_M_M_tau_NH} (Eqs.~\eqref{eq:CP_inv_U2_M_e_IH}-\eqref{eq:CP_inv_U2_M_M_mu_tau_IH}) for NH (IH) can be expressed as
\be
\label{eq:Yb_alpha_wHC_numbers}
 Y_B = - 3.2\times 10^{-20} \left( \frac{M}{\rm{GeV}} \right)^2 \left( \frac{1}{U^2} \right) f_\alpha^{\rm{H}} - 1.4 \times 10^{-28} \left( \frac{1}{U^2} \right)^2 \bar{f}_\alpha^{\rm{H}}\,.
\ee
The functions $f_{\alpha}^{\rm H}$ and $\bar{f}_\alpha^{\rm{H}}$ encode dependence of the CP invariants on the PMNS parameters.
$f_{\alpha}^{\rm H}$ for the relevant flavour can be obtained from Eq.~\eqref{eq:CP_inv_U2_M_e_NH} and Eq.~\eqref{eq:CP_inv_U2_M_mu_tau_IH} for NH and IH respectively:
\be
\label{eq:f_alpha}
f_e^{\rm{NH}} = -\frac{\sqrt{r}}{2} \theta_{13} s_{12} \sin(\delta + \phi)\,,  \,\,\,\,\,\,\, f_\mu^{\rm{IH}} = f_\tau^{\rm{IH}} = - \frac{r^2}{8} c_{12} s_{12} \sin\phi\,.
\ee
And $\bar{f}_\alpha^{\rm{H}}$ for the relevant flavour are obtained from Eq.~\eqref{eq:CP_inv_U2_M_M_e_NH}~(Eq.~\eqref{eq:CP_inv_U2_M_M_mu_tau_IH}) for NH (IH):
\bea
\label{eq:f_bar_alpha}
\bar{f}_e^{\rm{NH}} =   \frac{r}{2} c_{12}^2 \sin(2\phi)  \,,  \,\,\,\,\,\,\
 \bar{f}_\mu^{\rm{IH}}  = \bar{f}_\tau^{\rm{IH}}=  \frac{1}{2} \frac{r^2 c_{12}^2 s_{12}^2 \sin(2\phi)}{ 4 c_{12}^2 s_{12}^2 \cos^2(\phi) - 1 }\,.
\eea
This implies that successful baryogenesis represents a tight constraint on the CP violating phases $\delta$ and $\phi$.
For fixed $M$ and $U^2$ the CP phases are on one hand constrained to explain the observed asymmetry via Eq.~\eqref{eq:Yb_alpha_wHC_numbers} and on the other hand they are constrained by the condition $\Lambda_\alpha(x_{\rm{EW}}) \leq 1 \land \Lambda_\beta(x_{\rm{EW}}) \geq 1$.
The latter conditions requires to minimize $\epsilon_\alpha$. 
It is an interesting coincidence that for IH $\rm{min}(\eps_\alpha)$ is achieved for CP conserving values of the Dirac and Majorana phase. This leads to $f_\alpha^{\rm IH} ,\bar{f}_\alpha^{\rm IH}\to 0$ , see Fig.~\ref{fig:f_eps_fw_ov} and Eqs.~\eqref{eq:f_alpha} and ~\eqref{eq:f_bar_alpha}. 
The requirement of $Y_B > 0$ and $U^2 > 0$ leads to the following constraint
\be
\label{eq:f_wHC_constraint}
\left( \frac{M}{\rm{GeV}} \right)^4 \left( f_\alpha^{\rm{H}} \right)^2 \gtrsim 50  \bar{f}_\alpha^{\rm{H}}\,.
\ee 
This condition can be fulfilled in two ways, i) a significant suppression of $\bar{f}_\alpha^{\rm{H}}$ or ii) $\bar{f}_\alpha^{\rm{H}} < 0$
The former case corresponds to a significant tuning of the CP phases $(\delta, \phi)$, while the latter case is realizable in a much more extended phase space of $(\delta, \phi)$.
Therefore, for the masses of interest, $M = \mathcal{O}(1\, \mathrm{GeV})$, the asymmetry is most probably dominated by $\bar{f}_\alpha^{\mathrm{H}}$.
Hence, the mixing of the HNLs which can lead to successful baryogenesis is not only constrained by light neutrino data via Eq.~\eqref{eq:U2_fw}.
Rather, the value of the observed asymmetry sets a stronger constraint on $U^2$ as well as on the CP phases.

The upper bound on $U^2$ can be found as follows. 
For fixed $M$, increase $U^2$ as long as i) at least for some value of $(\delta, \phi)$ one flavour remains weak at $x_{\rm{EW}}$, that is $\Lambda_\alpha(x_{\rm EW}) <1$ (see Eq.~\eqref{eq:Gamma_alpha}), and ii) the very same phases are compatible with $ Y_B \geq Y_B^{\rm{obs}}$  (using Eq.~\eqref{eq:Yb_alpha_wHC_numbers}).
Additionally,  we demand that when two flavours $\alpha$ and $\beta$ are in different regimes $\Lambda_\alpha(x_{\mathrm{EW}}) < 1$ and $\Lambda_\beta(x_{\mathrm{EW}}) >1$,  they also satisfy  $\Lambda_\beta(x_{\mathrm{EW}}) - \Lambda_\alpha(x_{\mathrm{EW}}) > 1$. This ensures that the flavour $\beta$ is sufficiently strong with respect to $\alpha$  in order to guarantee the validity of the analytical expressions derived in Eqs.~\eqref{eq:ana_sol_cp_inv_fw_wlnv}-\eqref{eq:asym_fw_slnv}.
The so found upper bound is shown in Fig.~\ref{fig:scan}.

\noindent

\subsection{Regime $2$ -- flavoured sHC}
This regime is defined by having the helicity conserving interactions in thermal equilibrium, $\Lambda_M(x_{\mathrm{EW}}) \geq 1$, see Eq.~\eqref{eq:Gamma_slow_M_ov}.
This requires mixings of
\be
U^2 \geq 1 \times 10^{-6} \left( \frac{1\,\rm{GeV}}{M} \right)^{4}\,,
\ee
while being inside the range defined in Eq.~\eqref{eq:U2_fw}.
In Fig.~\ref{fig:regimes} this corresponds to the region above the blue dashed-dotted line and within the red dashed band.
Therefore, it is exclusively tested by FCC-ee.
The asymmetry is given by Eq.~\eqref{eq:asym_fw_slnv} and when including the CP invariants of Eqs.~\eqref{eq:CP_inv_U2_M_e_NH}-\eqref{eq:CP_inv_U2_M_mu_tau_NH} (Eqs.~\eqref{eq:CP_inv_U2_M_e_IH}-\eqref{eq:CP_inv_U2_M_mu_tau_IH}) for NH (IH) it can be expressed as
\be
\label{eq:Yb_alpha_sHC_numbers}
Y_B = -1.5\times10^{-25} \left( \frac{\rm{GeV}}{M}\right)^2 \left( \frac{1}{U^2} \right)^2 f_\alpha^{\rm{H}}\,.
\ee
The angular function $f_\alpha^{\rm{H}}$ is defined in Eq.~\eqref{eq:f_alpha}. 
Note that in this regime, for both IH and NH the limit $\eps_\alpha \to \rm{min}(\eps_\alpha)$ suppresses $Y_B$, since it implies that $f_\alpha^{\rm H} \to 0$ for the relevant slow flavours.
The upper bounds on the mixing compatible with baryogenesis can be found with the same strategy as outlined in the previous paragraph (see discussion after Eq.~\eqref{eq:f_wHC_constraint}) and are shown in Fig.~\ref{fig:scan}.



\section{Numerical results: parameter scan}
\label{sec:num}

\begin{figure}[!t]
\centering
\begin{tabular}{cc}
\hspace{-0.8cm}  \includegraphics[width=0.53\textwidth]{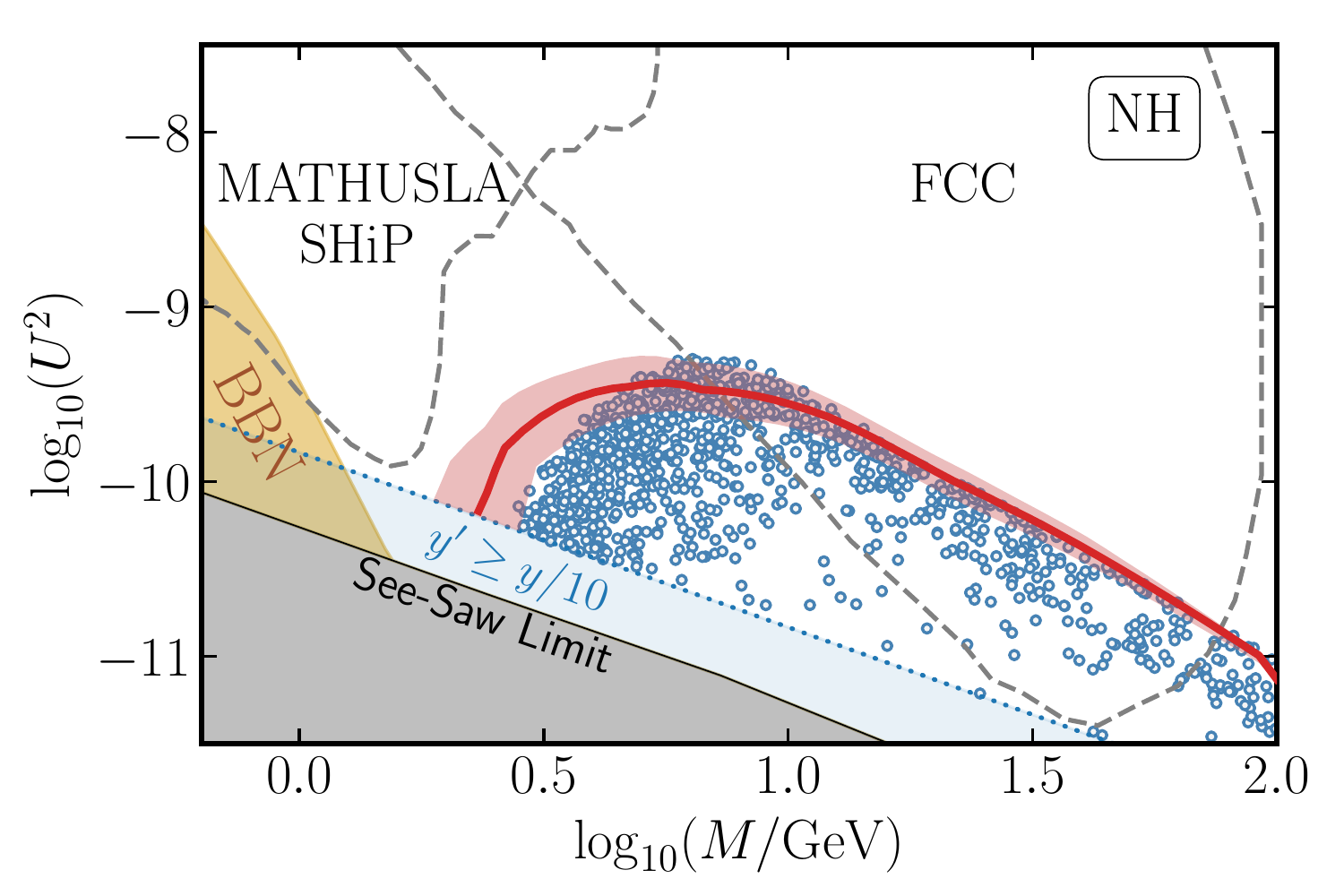} &
\hspace{-0.7cm}  \includegraphics[width=0.53\textwidth]{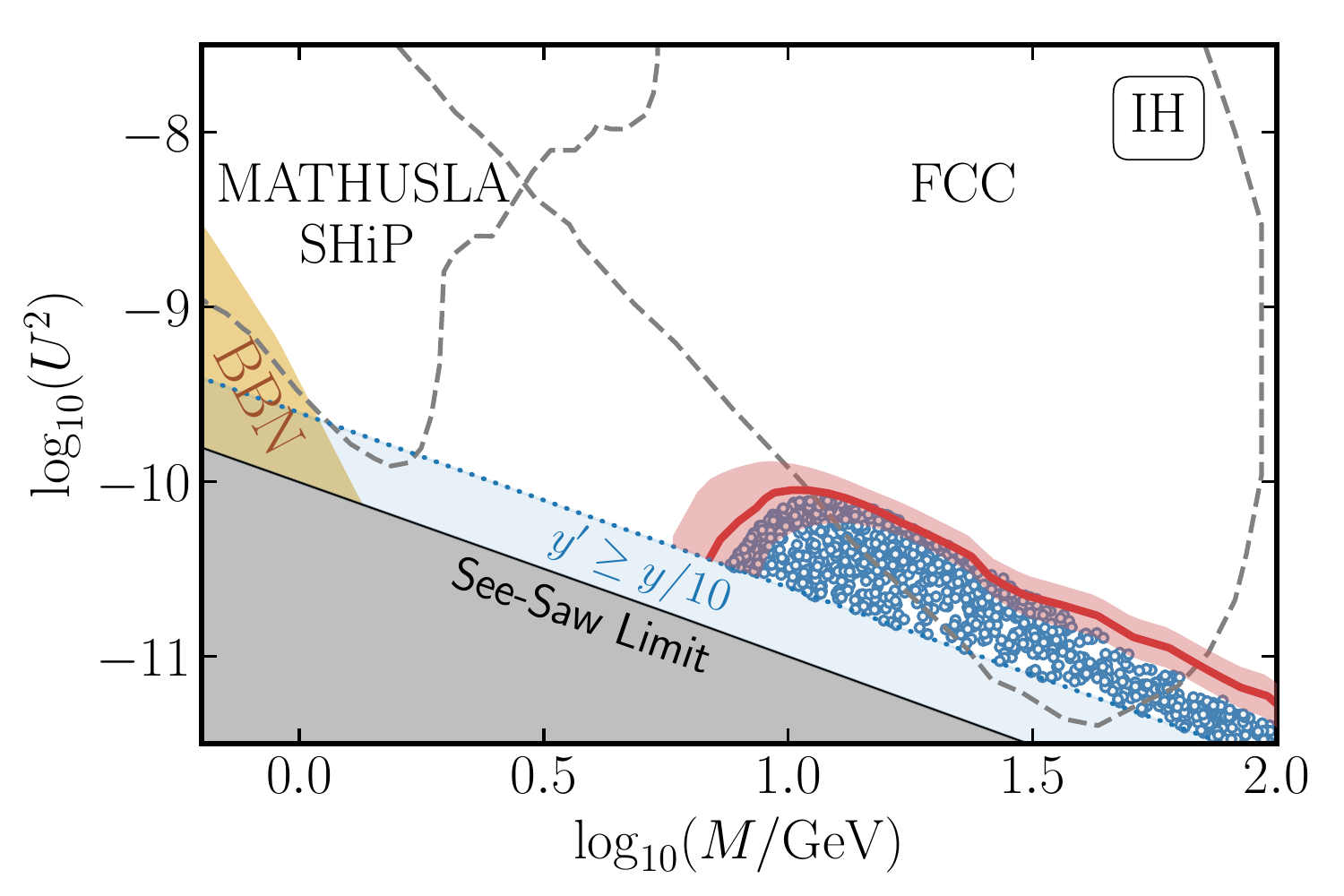} \\
\end{tabular}
\vspace{-0.4cm}
\caption{ 
Points leading to the observed baryon asymmetry for NH (IH) in the left (right) panel.
The red band indicates the analytical upper bound including a factor of two uncertainty in the final analytical result of $Y_B$.
}
\label{fig:scan}
\end{figure}

We numerically scan the parameter space to identify the regions in which the observed baryon asymmetry can be explained.
The analysis focuses on scenarios with \textit{exactly} degenerate HNLs.
Therefore, the complete parameter space is described by only four independent parameters that we choose as: the common mass $M$ of the HNLs, $y$ (or, equivalently, $U^2$) and the two PMNS CP phases $(\delta, \phi)$.
Future experiments as SHiP, MATHUSLA and FCC-ee will be sensitive to HNL masses in the range $0.3 \lesssim M/\rm{GeV} \lesssim 100$.
Because we are interested in identifying correlations between the baryon asymmetry and other observables, such as the masses and mixings of the HNLs, the numerical analysis focuses on this mass range.
We agnostically choose linearly flat priors in the two angles and logarithmically flat in $M$ and $y$.
The exact ranges are summarized in Tab.~\ref{tab:prior}.
\begin{table}[!t]
\begin{center}
\begin{tabular}{| c c c c |}
\hline
$\log_{10}(M/\rm{GeV})$ & $\log_{10}(y)$ & $\delta$ & $\phi$ \\
\hline
\hline
$[-1, 2]$ & $[-8, -4]$  & $[0, 2 \pi)$ & $[0, 2 \pi)$  \\
\hline
\end{tabular}
\caption{
Priors for the nested sampling.
}
\label{tab:prior}
\end{center}
\end{table}
Having set the prior volume, we can perform a nested sampling from the log-likelihood
\begin{align}
\log{\mathcal{L}} = -\frac{1}{2} \left( \frac{Y_B\left(T_{\rm{EW}}\right) - Y_B^{\rm{exp}}}{\sigma_{Y_B^{\rm{exp}}}} \right)\,.
\end{align}
We use amiqs~\cite{pilar_hernandez_2022_6866454}, a numerical software which we make publicly available on~\gitlink,  and implement it into the software UltraNest~\cite{2021JOSS....6.3001B}.
The result is shown in Fig.~\ref{fig:scan}.
%
%
%
In the same figure, the red line indicates the analytical upper bound on successful baryogenesis in the model obtained  in Sec.~\ref{subsec:parambounds_wHC}. The upper bound nicely reproduces the results of the scan when the approximate numerical solution (i.e. with constant rates) is used but  the agreement is slightly worse with the full numerical solution. 
The red band corresponds to a factor of two uncertainty in the value of $Y_B$ used to obtain the upper bound. This should account for the uncertainty arising from the varying interaction rates.
We further indicate in blue the region for which the perturbative expansion around the symmetric LN texture of Eq.~\eqref{eq:LNtexture} breaks down.
This only happens for the smallest mixings and is therefore not of particular interest to our analysis.

\begin{figure}[!t]
\centering
\begin{tabular}{cc}
\hspace{-0.5cm}  \includegraphics[width=0.49\textwidth]{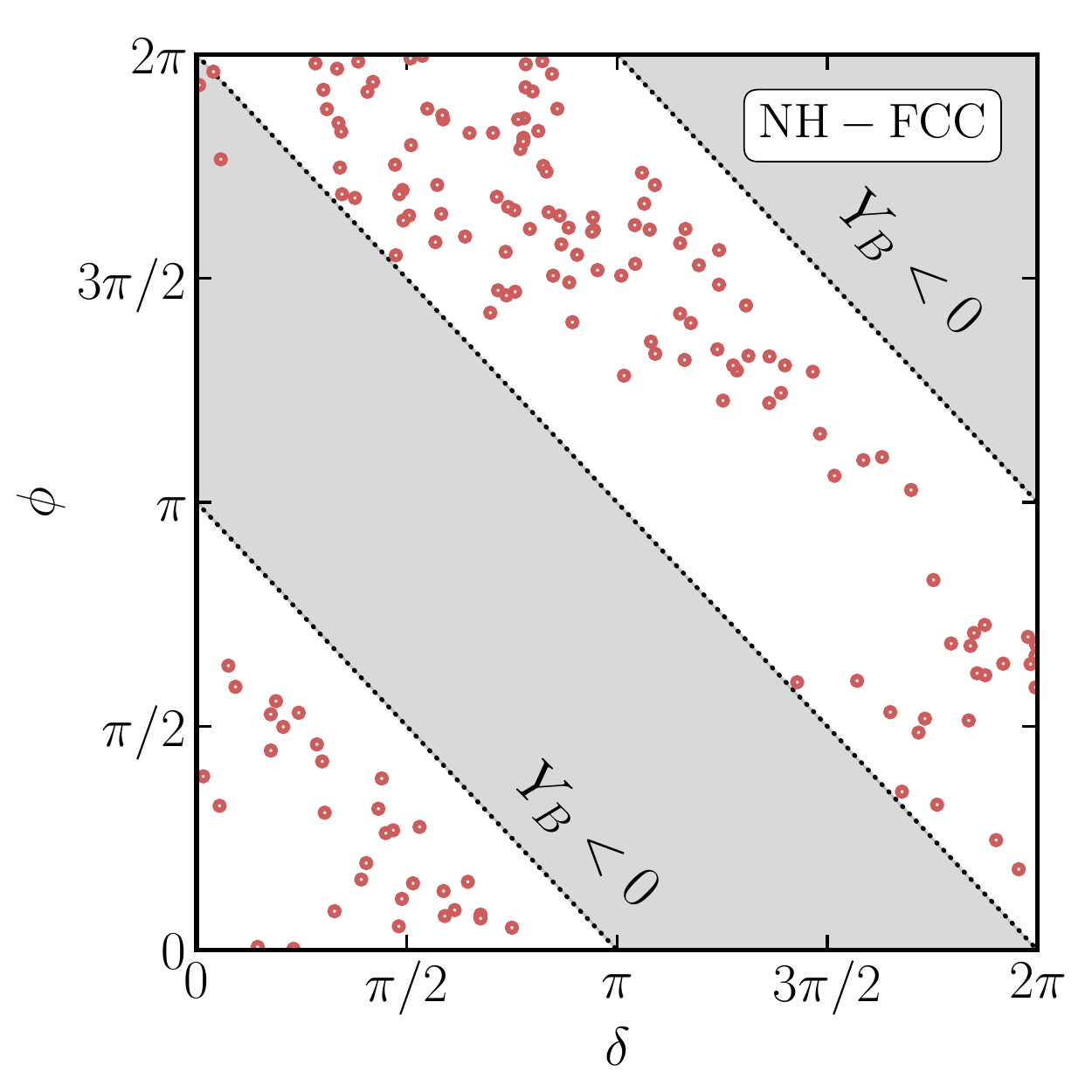} &
\hspace{-0.5cm}  \includegraphics[width=0.49\textwidth]{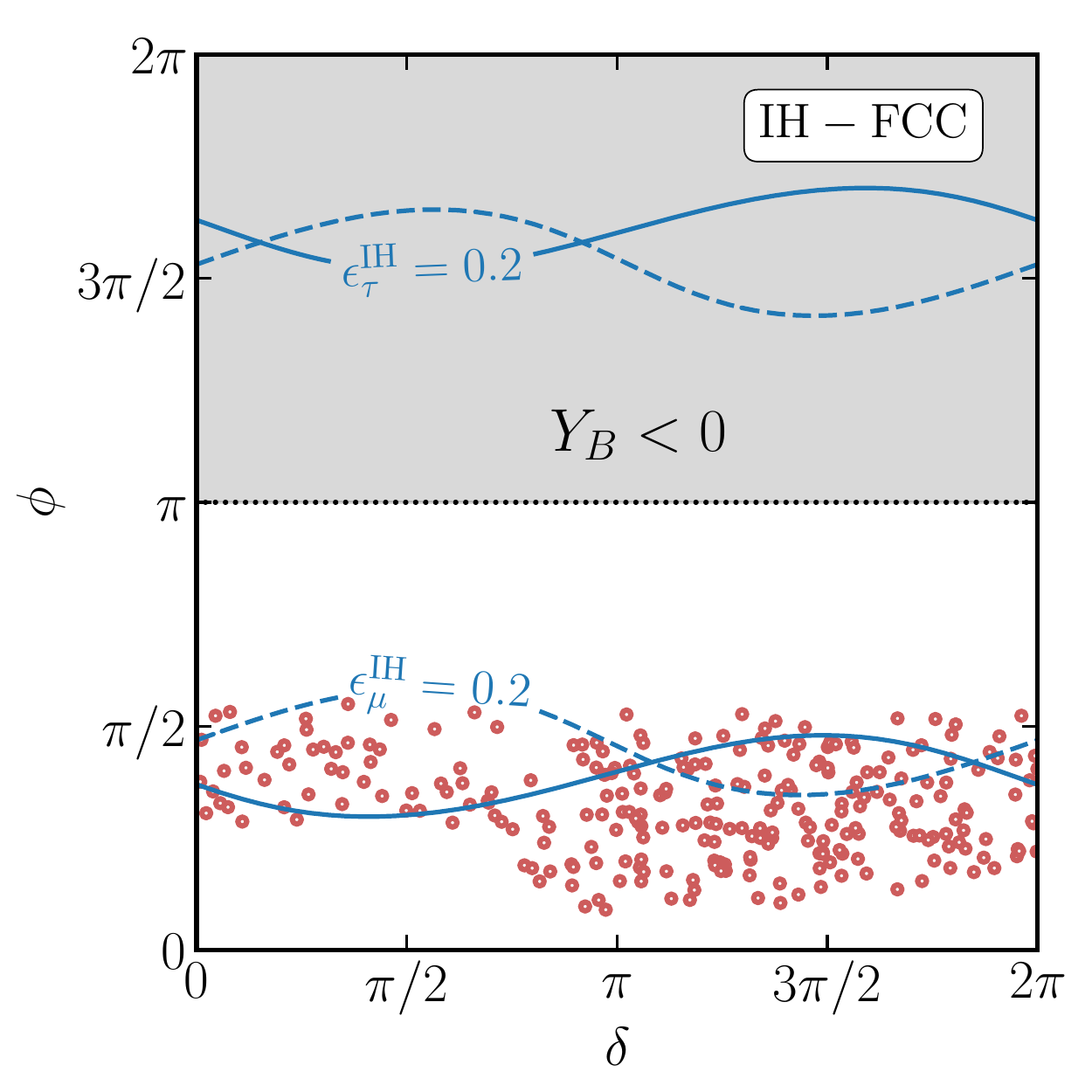} 
\end{tabular}
\vspace{-0.4cm}
\caption{ 
Points in the $(\delta, \phi)$ plane from a numerical scan leading to the observed baryon asymmetry for NH (IH) in the \textit{left} (\textit{right}) panel, with mixings in the sensitivity reach of FCC-ee.
\textit{Left}: The grey region is obtained by solving for ${f}^{\mathrm{NH}}_e < 0$, see Eq.~\eqref{eq:f_alpha} together with Eq.~\eqref{eq:Yb_alpha_sHC_numbers}. The slow flavour condition, $\epsilon_e \ll 1$, does not set any additional constraint.
\textit{Right}: The dashed and solid blue lines are obtained by solving for $\eps_{\mu} = 1/5$ and $\eps_\tau = 1/5$, respectively. The grey region leads to the wrong sign in the baryon asymmetry derived from Eq.~\eqref{eq:f_alpha} together with Eq.~\eqref{eq:Yb_alpha_sHC_numbers}.
}
\label{fig:scan_deltaphi}
\end{figure}

We see that only FCC-ee is sensitive to the parameter space which can explain the observed baryon asymmetry in the model.
In particular, only the regime with strong helicity conserving interactions will be probed. 
Note that in this regime our analytical solutions are particularly robust, as discussed in Sec.~\ref{subsec:ana_vs_num}. 
The analytical and numerical results show that the parameter space for successful baryogenesis in the model is far more restricted than naively expected from the CP invariants of Eqs.~\eqref{eq:cpinv_flavoured_param}-\eqref{eq:cpinv_unflavoured_weak_param}. 
In particular, the generated asymmetry is given by Eq.~\eqref{eq:asym_fw_slnv}, which only depends on the CP invariant of Eq.~\eqref{eq:cpinv_flavoured_param}.
This leads to the parametrical dependence on the CP violating phases as given in Eq.~\eqref{eq:f_alpha}, which differ in NH and IH.
The constraints on the phases are shown in Fig.~\ref{fig:scan_deltaphi}.
%
%
\begin{figure}[!t]
\centering
\includegraphics[width=0.55 \textwidth]{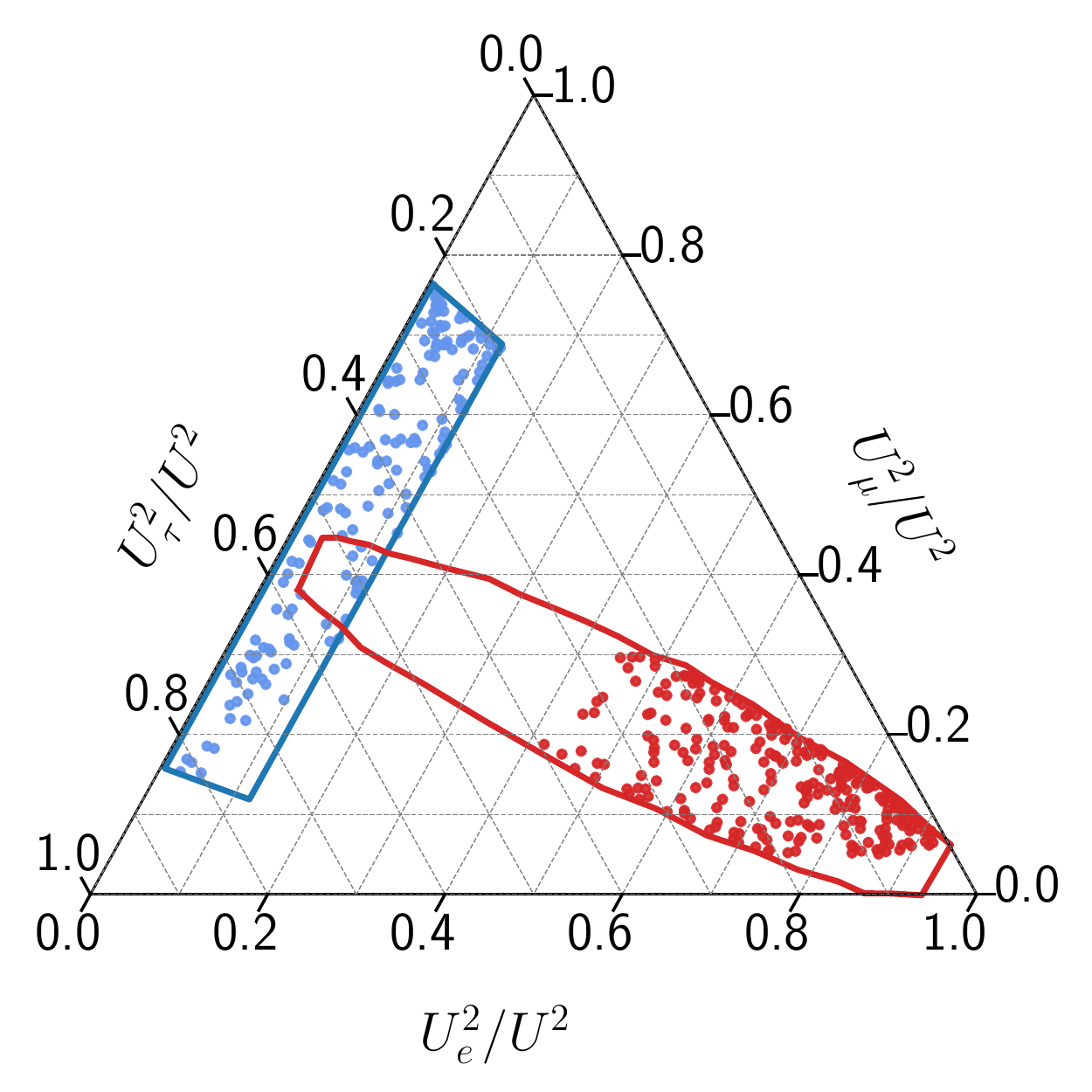} 
\vspace{-0.4cm}
\caption{ 
Ternary plot of points from a numerical scan leading to the observed baryon asymmetry for NH (IH) in blue (red), with mixings in the sensitivity reach of FCC-ee.
The enclosed region is the allowed region by neutrino oscillations data.
}
\label{fig:ternary}
\end{figure}
%
%
On the other hand, the constraints on the phases also emerge as constraints on the flavour dependent mixings of the HNLs, which can be parametrized by 
\begin{align}
\frac{U_\alpha^2}{U^2} \equiv \frac{1}{2}\sum_{I}\frac{|\Theta_{\alpha I}|^2}{U^2}\simeq \epsilon_\alpha \,.
\end{align}
The result is shown in the ternary plot of Fig.~\ref{fig:ternary}.
Especially in the IH scenario the constraint from the baryon asymmetry on the flavour ratio is apparent.
The requirement $\phi < \pi$, see Eq.~\eqref{eq:Yb_alpha_sHC_numbers} together with Eq.~\eqref{eq:f_alpha}, selects $|U_e^2| \gg |U^2_{\mu,\tau}|$.

We note that the  baryon asymmetry depends strongly on the exact HNL interaction rates and therefore uncertainties in the latter induce a large uncertainty in the constraints on HNL (flavour) mixings derived from successful baryogenesis.



\section{Predicting the baryon asymmetry}
\label{sec:potential_measure}
\begin{figure}[!t]
\centering
\begin{tabular}{cc}
\hspace{-0.5cm}  \includegraphics[width=0.49\textwidth]{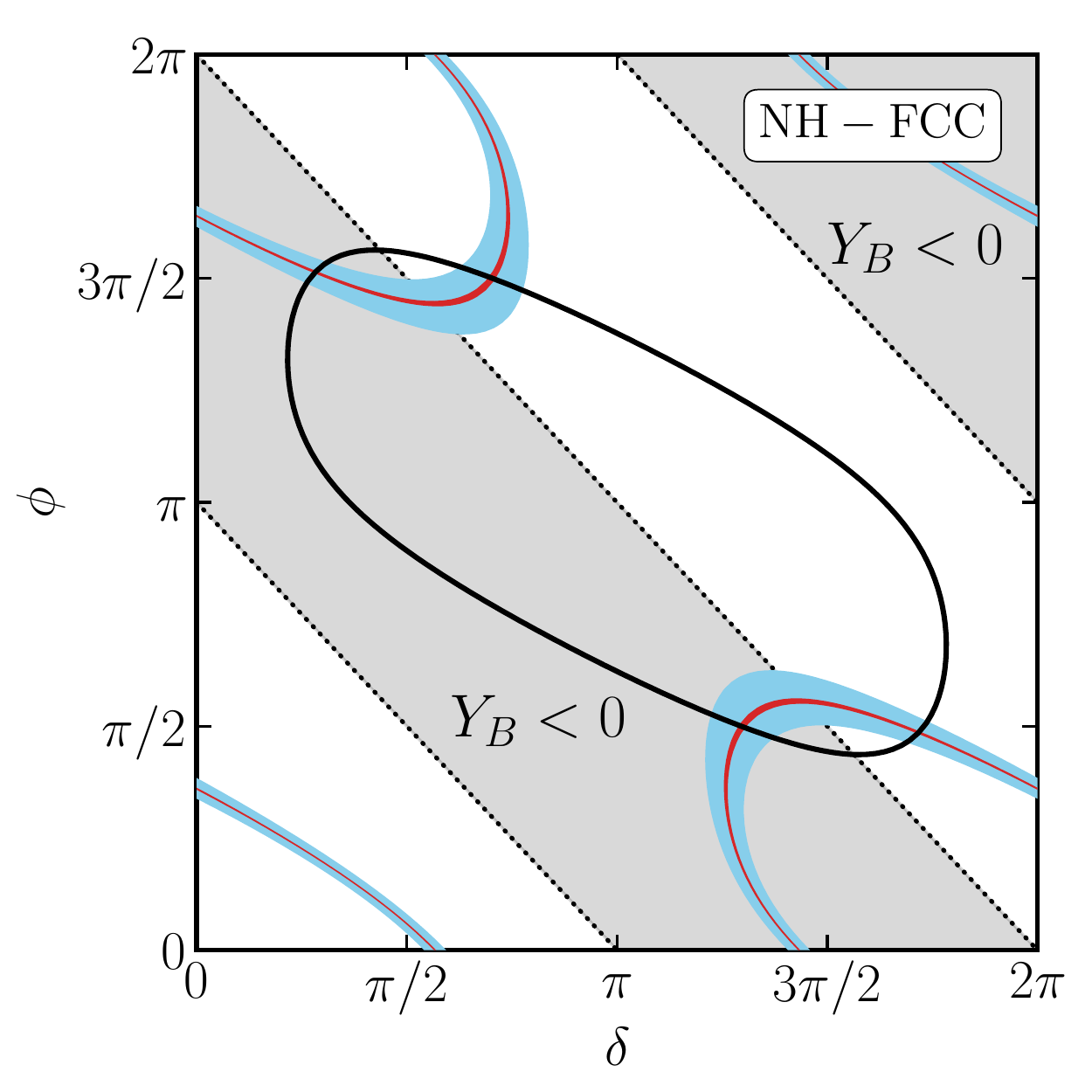} &
\hspace{-0.5cm}  \includegraphics[width=0.49\textwidth]{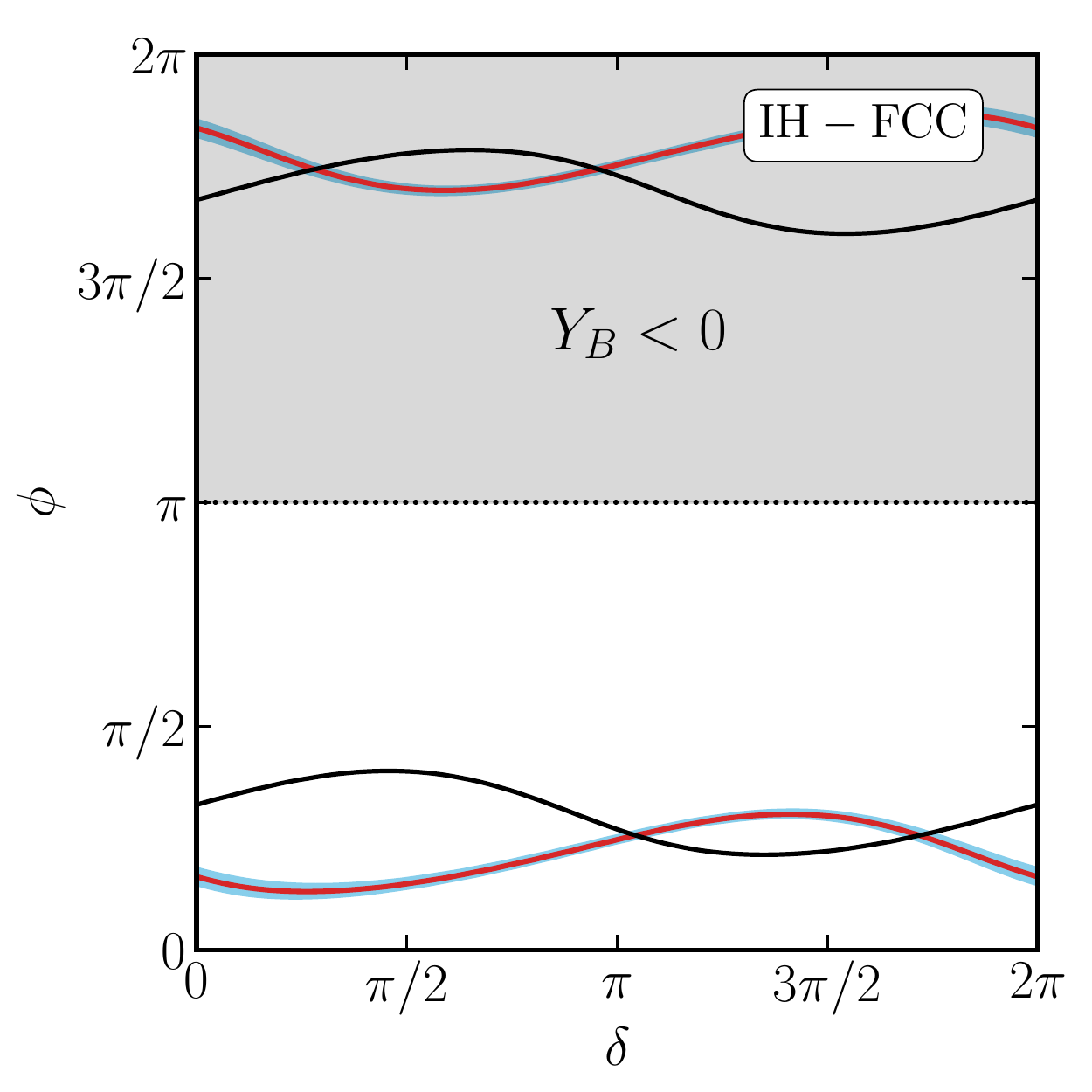} 
\end{tabular}
\vspace{-0.4cm}
\caption{ 
Isocurvature lines of Eqs.~\eqref{eq:Ue_eps_NH}-\eqref{eq:Utau_eps_IH} for an assumed flavour dependent mixing measured as given in Tab.~\ref{tab:potential_measure}.
NH is shown on the left and IH on the right panel.
The black (red) lines represent the constraints from a measurement of $U_e^2/U_\mu^2$($U_e^2/U_\tau^2$) with a 1$\%$ error. The blue band represents the constraints from a measurement of  $U_e^2/U_\tau^2$ with a $10\,\%$ uncertainty.
The grey regions lead to the wrong sign in the asymmetry generations, see Eq.~\eqref{eq:f_alpha}.
}
\label{fig:delta_phi_iso}
\end{figure}

\begin{figure}[!t]
\centering
\begin{tabular}{cc}
\hspace{-0.5cm}  \includegraphics[width=0.49\textwidth]{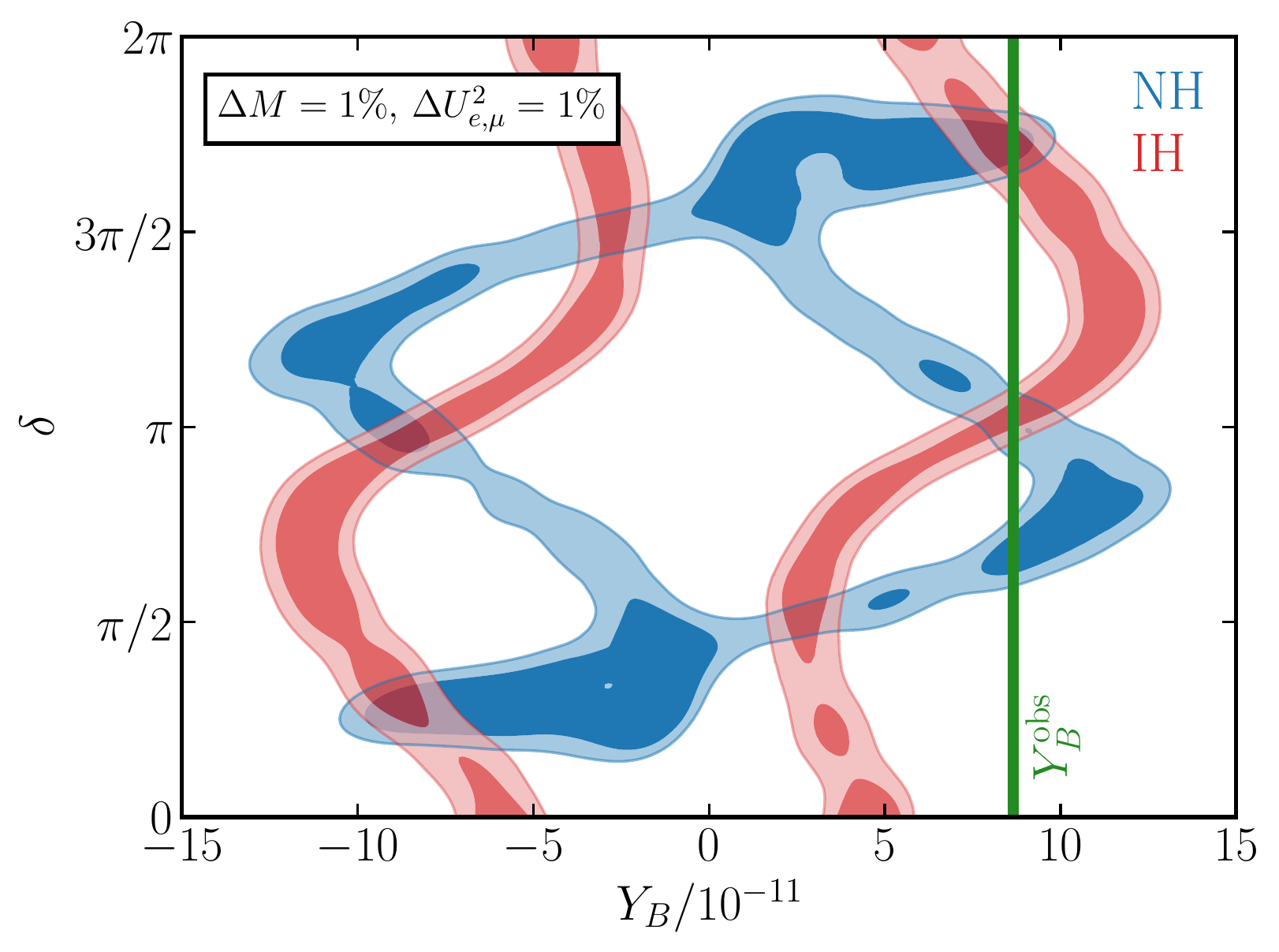} &
\hspace{-0.5cm}  \includegraphics[width=0.49\textwidth]{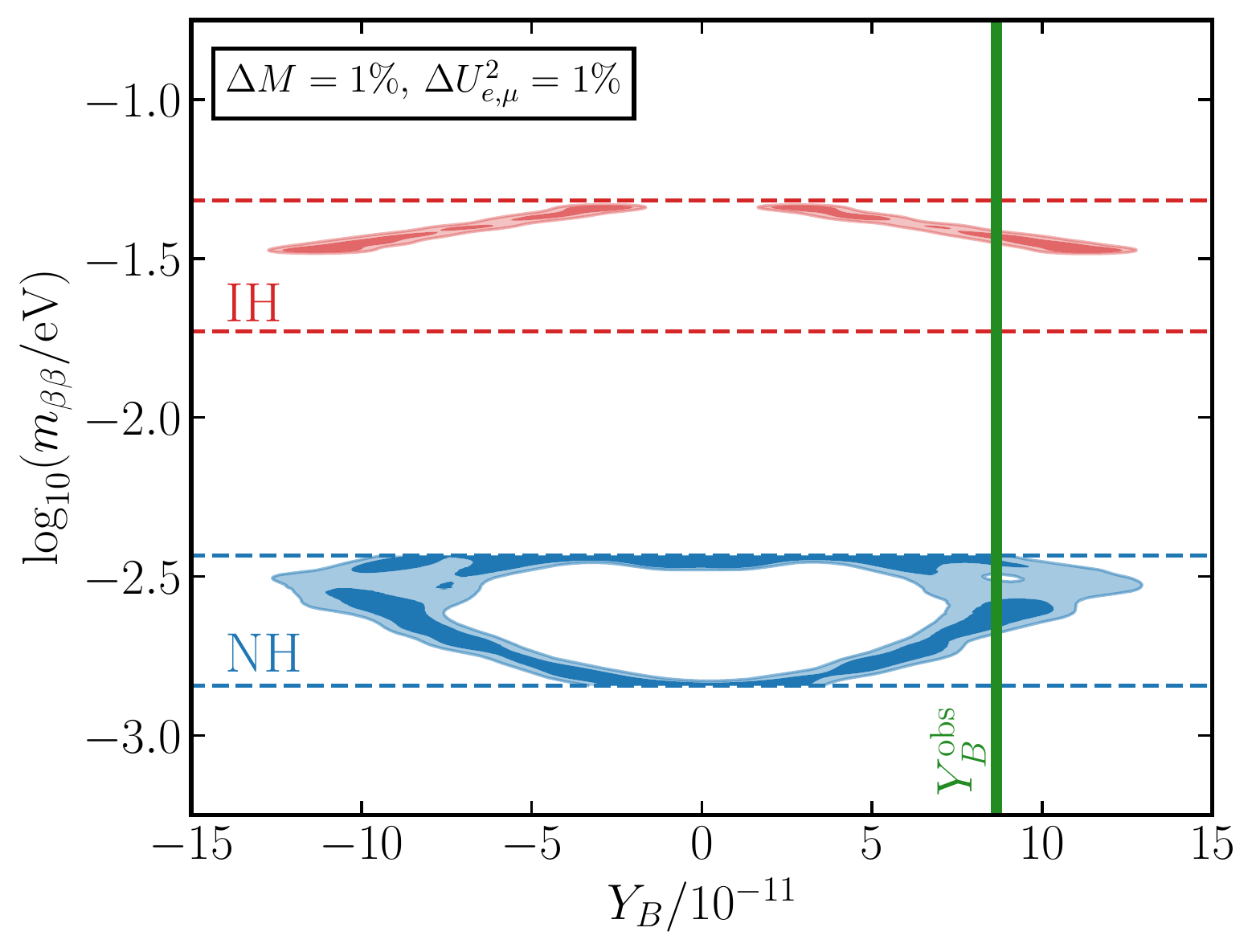} \\
\hspace{-0.5cm} \includegraphics[width=0.49\textwidth]{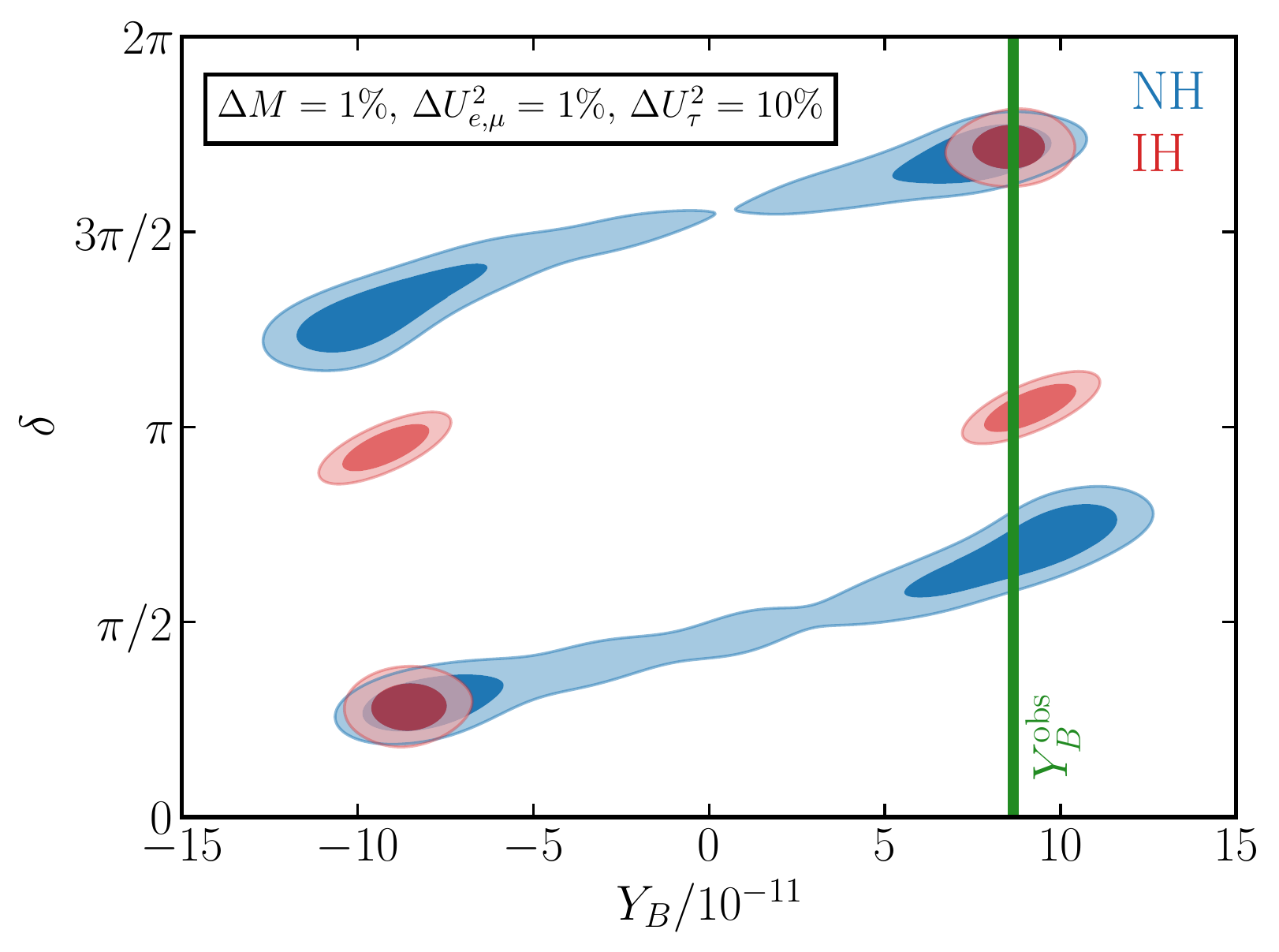}  &
\hspace{-0.5cm} \includegraphics[width=0.49\textwidth]{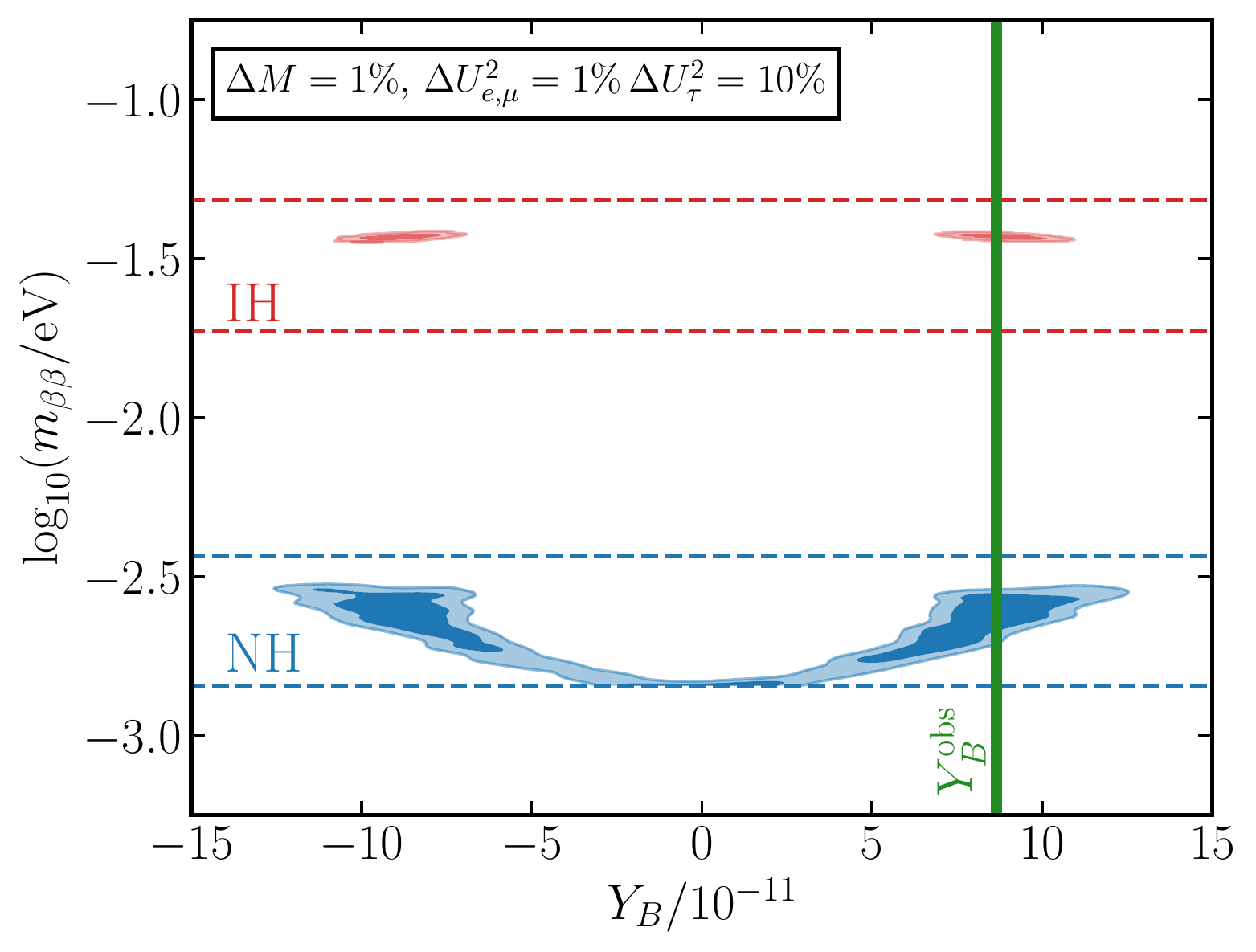}  \\
\hspace{-0.5cm} \includegraphics[width=0.49\textwidth]{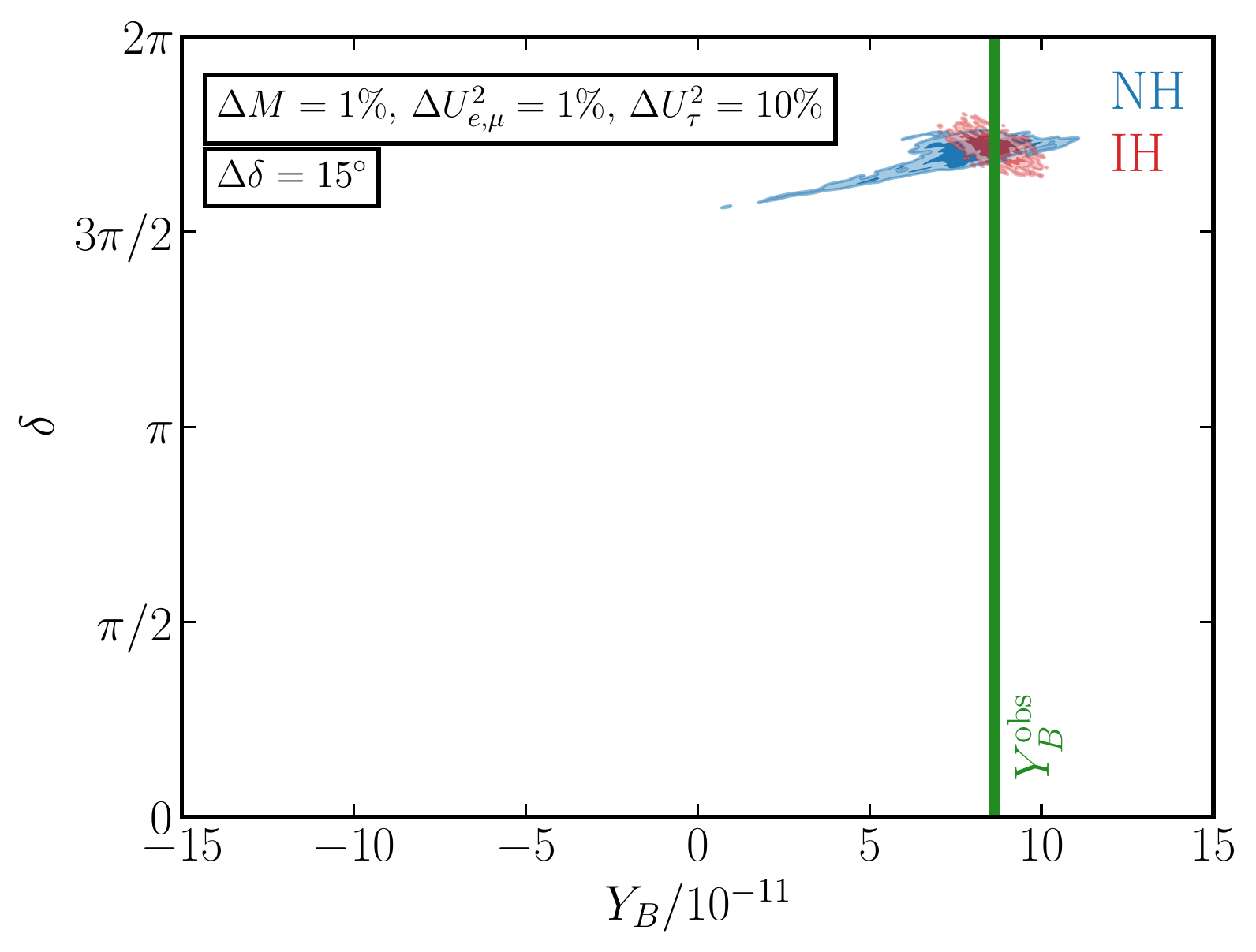}  &
\hspace{-0.5cm} \includegraphics[width=0.49\textwidth]{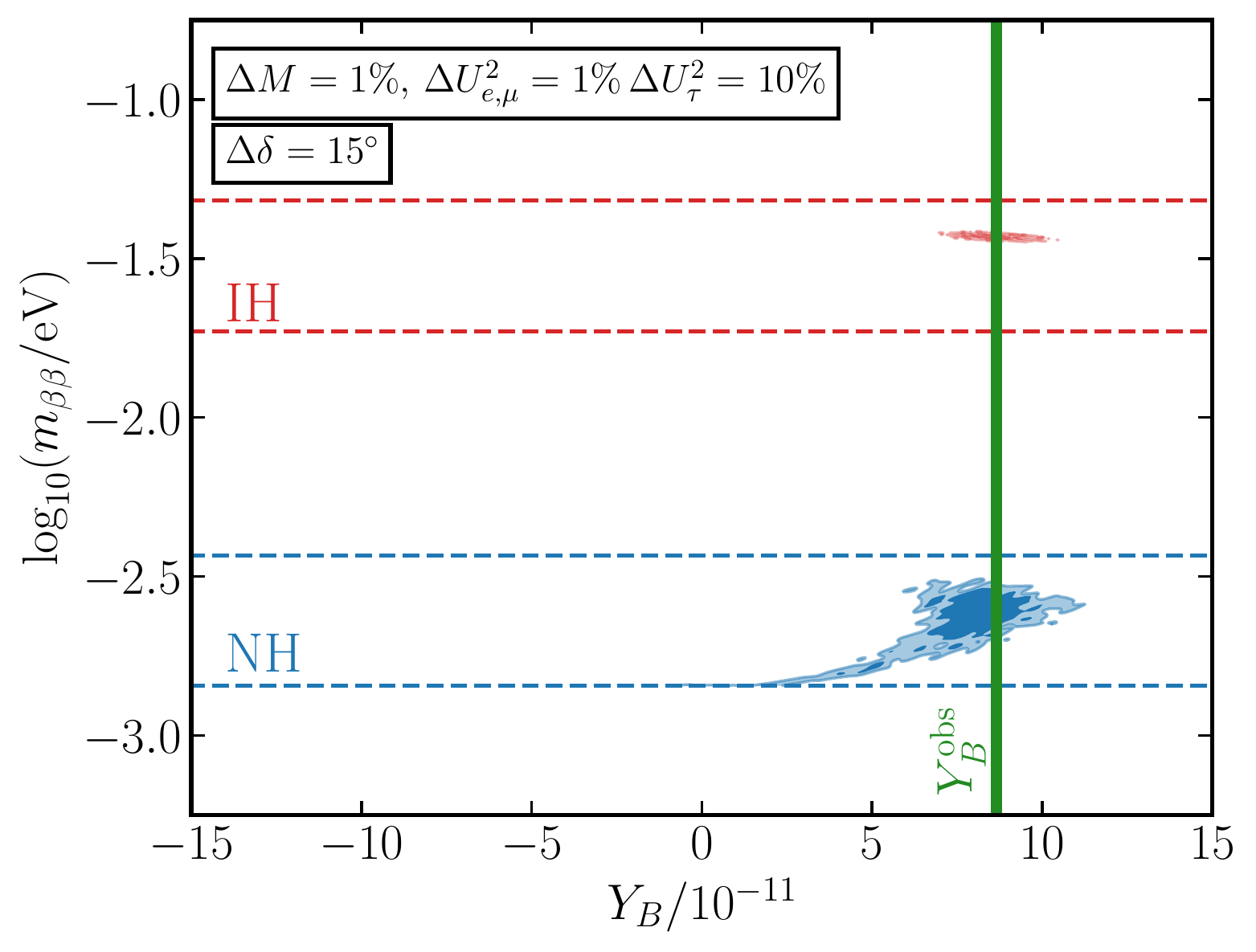}  \\
\end{tabular}
\vspace{-0.4cm}
\caption{ 
Result of a numerical likelihood inference in the case of a measurement of $(M, U_e^2, U_\mu^2)$ (top), $(M, U_e^2, U_\mu^2, U_\tau^2)$ (middle) and $(M, U_e^2, U_\mu^2, U_\tau^2, \delta)$ (bottom), with the true values and errors given in Tab.~\ref{tab:potential_measure}.
All plots show the correlation between $\delta$ (left panel) and $m_{\beta\beta}$ (right panel) with the baryon asymmetry.
The blue regions represent the NH case and the red regions the IH.
The green vertical line represents the observed value of the asymmetry.
The red and blue dashed horizontal lines in the right panel indicate the range of $m_{\beta\beta}$ predicted by the model with current neutrino oscillations data. 
}
\label{fig:reconstruction}
\end{figure}

In this section we show how laboratory measurements at future colliders and neutrino oscillations facilities can be used to predict the baryon asymmetry and the neutrinoless double-beta decay rate.
From the analytical formulae for the baryon asymmetry we see that it can break the degeneracies in the CP violating phases that remain from measuring the HNL mass and its flavour mixings.
Furthermore, when we include the measurement of the Dirac CP phase, all the free parameters of the model can be determined.
In the following and for the rest of the section we assume two benchmark points for NH and IH that reproduce the baryon asymmetry and lead to the measurable neutrino parameters in Tab.~\ref{tab:potential_measure}. 
%
\begin{table}[!t]
\begin{center}
\begin{tabular}{| c | c c c c | c |}
\hline
 & $M^{\rm true}/\rm{GeV}$ & $(U_e^2)_{\rm true}$ & $(U_\mu^2)_{\rm true}$ & $(U_\tau^2)_{\rm true}$ & $\delta^{\rm true}/\rm{rad}$ \\
\hline
\hline
NH & $31.60$ & $2.843\times 10^{-12}$  & $ 1.087 \times 10^{-11} $ & $1.234 \times 10^{-11}  $ & $ 5.396 $\\
IH & $20.731$ & $3.291\times 10^{-11}$  & $ 4.823 \times 10^{-12} $ & $3.465 \times 10^{-12}  $ & $5.402 $  \\
\hline
\end{tabular}
\caption{
Measurable neutrino parameters for two benchmark points that reproduce the baryon asymmetry. 
}
\label{tab:potential_measure}
\end{center}
\end{table}
%
%
%
\vspace{0.25 cm}
\newline
\noindent
\textbf{Measurement of $M$ and $U_\alpha^2$.}
The exact expressions for the flavour dependent HNL mixings are somewhat lengthy.
We can get accurate simpler expressions by expanding  in the small parameters
\be
r \equiv \frac{\sqrt{\Delta m_{\rm sol}^2}}{\sqrt{\Delta m_{\rm atm}^2}} \sim \theta_{13} \sim \delta_{23} \equiv \theta_{23} - \pi/4 \sim y'/y \sim 10^{-1}\,.
\ee
The result is
\bea
\label{eq:Ue_eps_NH}
U_{e}^2/U^2 &\simeq& 2 \sqrt{r} \theta_{13} s_{12}  \cos\left( \delta + \phi \right) + r s^2_{12}  \,,\\
U_{\mu}^2/U^2 &\simeq& \frac{1}{2}\left[1 + 2\delta_{23}  - r s^2_{12} + 2 \sqrt{r} c_{12} \cos\phi   \right]\,,\\
U_{\tau}^2/U^2 &\simeq& \frac{1}{2} \left[ 1 -  2\delta_{23}  - r s^2_{12} - 2 \sqrt{r} c_{12} \cos\phi \right]\,,
\eea
for NH and 
\bea
U_{e}^2/U^2 &\simeq& \frac{1}{2} \left[ 1 + \sin(2\theta_{12}) \cos\phi \right]\,,\\
\nonumber
U_{\mu}^2/U^2 &\simeq&  \frac{1}{4} \left[  (1 - 2 \delta_{23})(1 - \sin2\theta_{12}\cos\phi) - 2 \theta_{13} \sin\delta \sin\phi  \right.  \\
&-& \left. 2\cos2\theta_{12}\theta_{13} \cos\delta \cos\phi \right] \,,\\
\label{eq:Utau_eps_IH}
\nonumber
U_{\tau}^2/U^2 &\simeq&  \frac{1}{4} \left[ (1 + 2\delta_{23})(1 - \sin2\theta_{12}\cos\phi) + 2\theta_{13} \sin\delta \sin\phi  \right.  \\
&+& \left. 2 \cos2\theta_{12} \theta_{13} \cos\delta \cos\phi \right] \,,
\eea
for IH. 
It is clear from these expressions that a measurement of the HNL mass and its flavour dependent mixings can constrain both, the Dirac $\delta$ and Majorana $\phi$, CP violating phases of the model \cite{Hernandez:2016kel,Caputo:2016ojx}.
This can be visualized with isocurvature lines of $U_e^2/U_\beta^2$, with $\beta = \{\mu, \tau\}$, projected on the plane of $(\delta, \phi)$, as shown in Fig.~\ref{fig:delta_phi_iso}.
%
%
%
%
We assume a $1\,\%$ error in the determination of $U_{e,\mu}^2$ and a $10\,\%$ error for $U_\tau^2$.
We see that only a measurement of all three individual flavour mixings can significantly constraint both CP phases.
In particular, for NH and IH such a measurement can predict $(\delta, \phi)$ up to a fourfold degeneracy.
Demanding that the observed baryon asymmetry should be explained reduces the degeneracy and predicts two values of each CP phase.
On the other hand, the Dirac CP phase will be measured at neutrino oscillations facilites which then will point to a unique value of the Majorana CP phase predicted by the model.
\vspace{0.25 cm}
\newline
\noindent
\textbf{Measurement of $M,\, U_\alpha^2$ and $\delta$.}
We add now to the previous assumption concerning the measurement of the HNL mass and mixings the measurement of the $\delta$ phase. The expected uncertainty is $\delta$ depends on the true value of the parameter. For the value in Tab.~\ref{tab:potential_measure}, we assume the conservative estimate of a relative precision of $15\degree$ for a run time of $10$ years~\cite{DUNE:2020ypp}. 
No additional measurement is in principle necessary to fix the Majorana phase $\phi$,  the baryon asymmetry and the amplitude for neutrinoless double-beta decay. Conversely, a putative measurement of neutrinoless double-beta decay  can add important constraints on $\phi$. 
The amplitude for the latter process depends on the combination of neutrino parameters $m_{\beta\beta}$, that in the model considered is well approximated by
\begin{equation}
m_{\beta\beta}=\left| \sum _{i={\rm light}} \left(U_{\nu}\right)^2_{ei}m_i \right|\,,
\end{equation}
In Fig.~\ref{fig:reconstruction} we show the prediction for $m_{\beta\beta}$ and $Y_B$ from putative measurements of i) $(M, U_e^2, U_\mu^2)$, ii) $(M, U_e^2, U_\mu^2, U_\tau^2)$ and iii) $(M, U_e^2, U_\mu^2, U_\tau^2, \delta)$, with the assumed experimental uncertainties indicated in the plot.\footnote{The error on the matrix elements for neutrinoless double-beta decay has not been included.}
%
%
%
%
%
We see how the collection of such measurements predict both $m_{\beta\beta}$ and $Y_B$. 
\vspace{0.25 cm}
\newline
\noindent
\textbf{Measurement of HNL oscillations.}
In the minimal model considered here the heavy neutrinos are completely degenerate before the EWPT. A mass splitting proportional to the Yukawa couplings is generated after EWPT when the Higgs develops its vacuum expectation value. In particular, the HNL splitting is given by~\cite{Ibarra:2010xw,Antusch:2017ebe,Penedo:2017knr,Drewes:2019byd,Shaposhnikov:2008pf}
\bea
\label{eq:dMcorrectionsNH}
\Delta M_{\rm{NH}}&=&|m_3|-|m_2|=\sqrt{\Delta m^2_{\rm{atm}}}-\sqrt{\Delta m^2_{\rm{sol}}},\\
\label{eq:dMcorrectionsIH}
\Delta M_{\rm{IH}}&=&|m_2|-|m_1|=\sqrt{\Delta m^2_{\rm{atm}}}-\sqrt{\Delta m^2_{\rm{atm}}-\Delta m^2_{\rm{sol}}},
\eea
up to one loop corrections. In principle these small splittings are in the right ballpark for the observation of HNL oscillations, which roughly requires $\Gamma\sim \Delta M$. However, imposing this condition, we find that the region of the parameter space in which this is fulfilled lies above the region of the $U^2$ vs $M$ plain in which the baryon asymmetry can be successfully generated. This means that a signal associated to HNL oscillations in colliders would exclude this minimal mechanism for the baryon asymmetry generation. Conversely,  $\Delta M> \Gamma$ is satisfied for the part of the parameter space compatible with the observed baryon asymmetry and, thus, a LN-violating signal from the HNL decay would be potentially observed at FCC-ee~\cite{Hernandez:2018cgc,Drewes:2019byd}. The measurement of such a signal would provide a powerful test of the scenario and add complementary information to the measurements already considered above.

\begin{figure}[!t]
\centering
\begin{tabular}{cc}
\hspace{-0.8cm}  \includegraphics[width=0.53\textwidth]{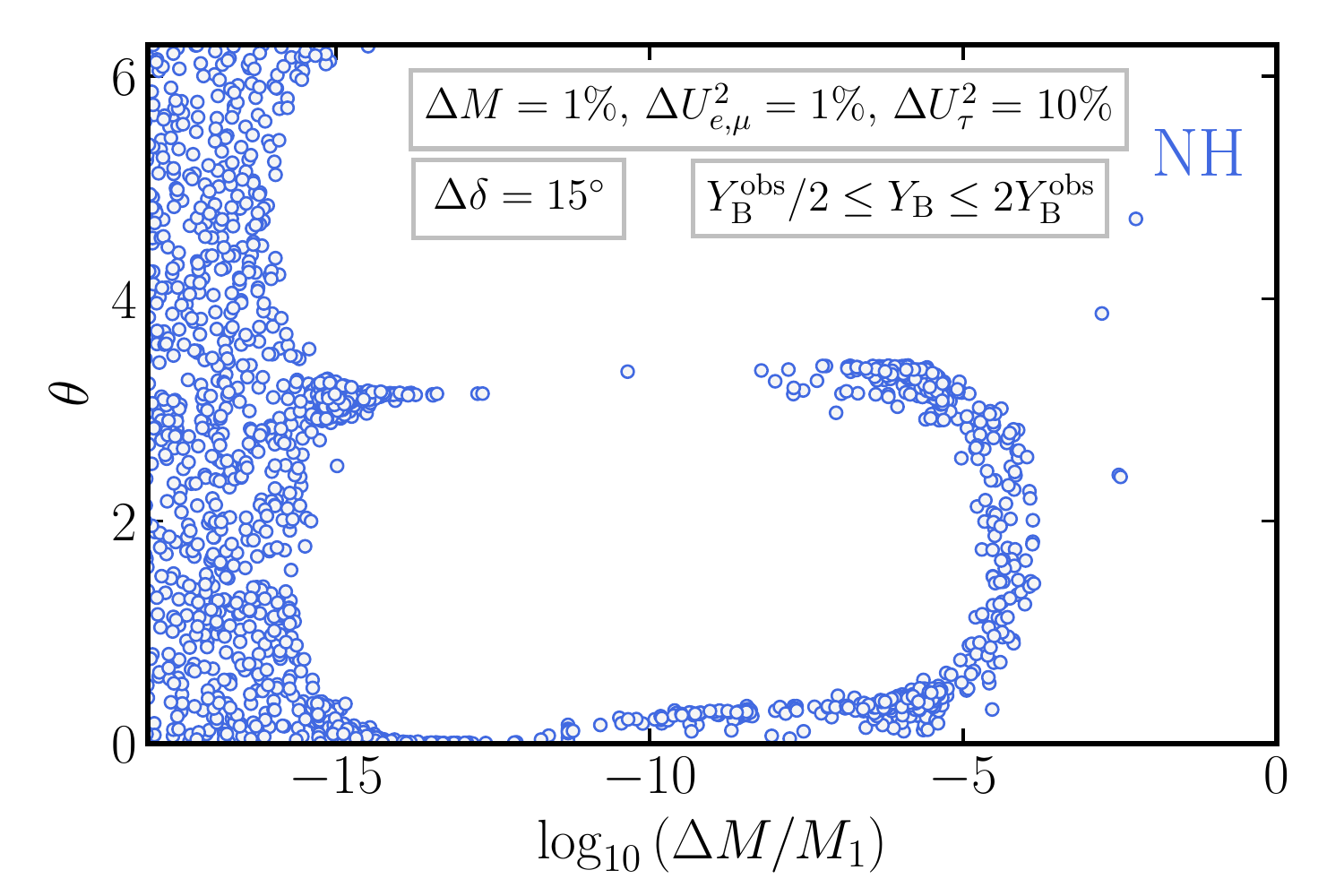} &
\hspace{-0.7cm}  \includegraphics[width=0.53\textwidth]{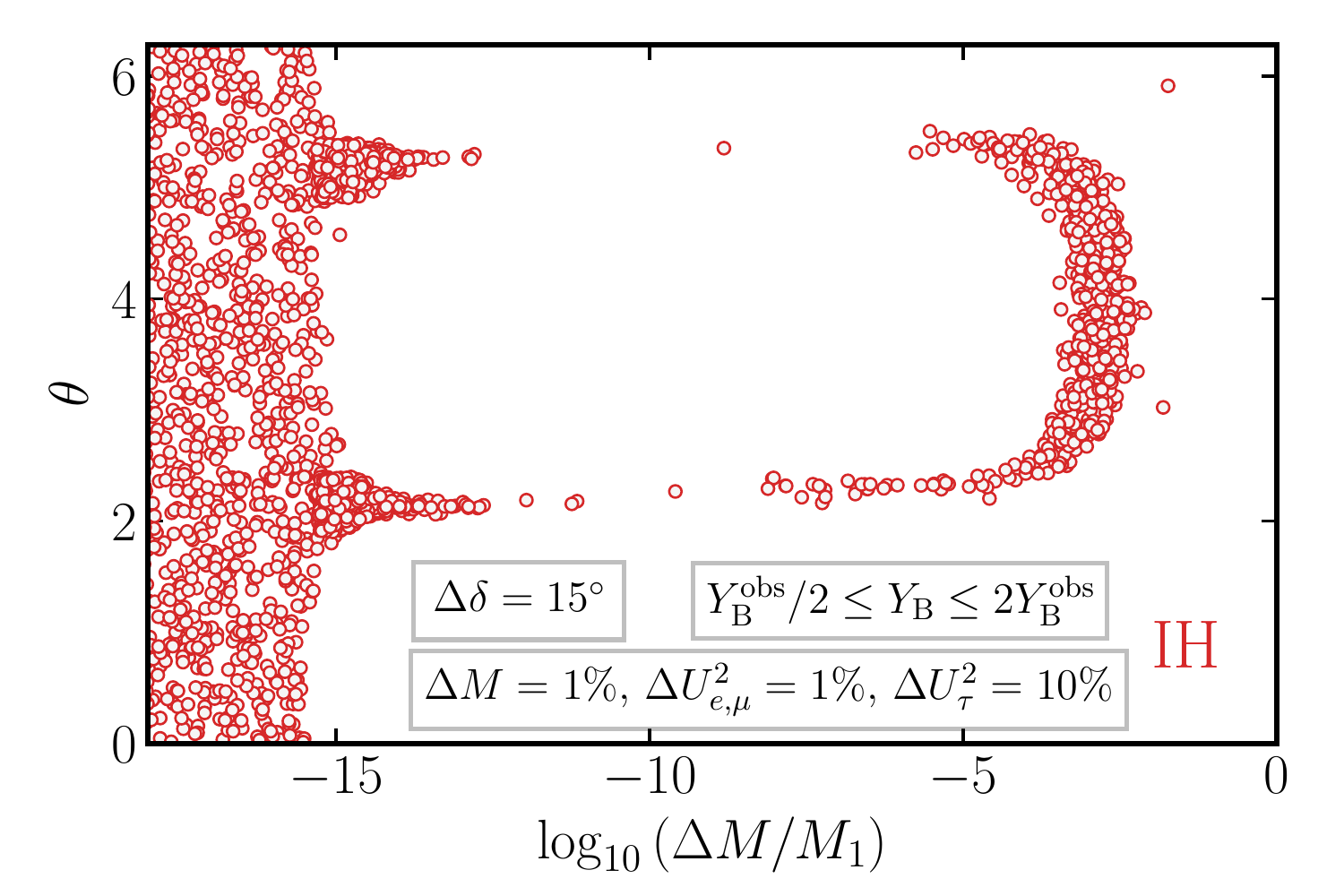} \\
\end{tabular}
\vspace{-0.4cm}
\caption{ 
Points leading to the observed baryon asymmetry in the non-degenerate scenario for NH (left panel) and IH (right panel) assuming the same putative measurement of the neutrino parameters as in the bottom panel of Fig.~\ref{fig:reconstruction}. 
}
\label{fig:dM0vsdM}
\end{figure}

\textbf{Degenerate vs non-degenerate case.}
As we have seen above, a measurement of HNL oscillations would exclude this minimal model as the origin of the baryon asymmetry. 
An interesting related question is whether it is experimentally possible to discern the model with exactly degenerate HNLs from the general case with a non-zero HNL mass splitting, if both successfully generate the baryon asymmetry. 
Note that once the HNLs are non-degenerate, besides their mass splitting $\Delta M$, there is an additional degree of freedom in the theory: 
a new CP-violating phase $\theta$ appears, which is associated to the breaking of LN in the heavy sector.
In order to understand this issue, we have considered two benchmark points for NH and IH that lead to the measured parameters in Tab.~\ref{tab:potential_measure} with the same precision considered above. 
Scanning over the six free parameters of the non-degenerate model, we show in Fig.~\ref{fig:dM0vsdM} the projection of the points in the plane $(\Delta M/M_1, \theta)$ which successfully explain the observed value of the baryon asymmetry.
Three observations are in order. 
First, for values of $\Delta M/M_1 \lesssim 5\times 10^{-16}$ the baryon asymmetry becomes independent of the high-scale phase $\theta$.
This indicates that the generation of the baryon asymmetry is dominated by the thermally induced HNL mass splitting in this range and 
the results derived within the degenerate limit apply. 
Secondly, another densely populated region with HNL mass splittings $\Delta M/M_1 \simeq 10^{-4}\,$ $(10^{-3})$ for NH (IH) can explain the observed baryon asymmetry for any $\theta$ within the range $0\leq \theta \leq \pi$ $(2.1 \leq \theta \leq 5.2)$ for NH (IH).
Values of $\theta$ outside of this range will lead to a wrong sign in the baryon asymmetry, see  Ref.~\cite{Hernandez:2022ivz}.
Finally, for mass splittings in between these two limits, the asymmetry can be generated only for fine-tuned values of $\theta$. This happens at the limit of the previously mentioned ranges of $\theta$. 
The reason is the following.
If a set of parameters $(M_1, y, \delta, \phi)$ successfully explains the baryon asymmetry with degenerate HNLs, then typically the same set of parameters in the non-degenerate case significantly overshoots the asymmetry.
This overshooting can only be compensated by adjusting $\theta$ in order to suppress the corresponding CP invariants, see Ref.~\cite{Hernandez:2022ivz}.  Within the assumed experimental accuracy, a measurement of $(M, U^2_\alpha, \delta)$ is therefore not sufficient to disentangle whether the baryon asymmetry is produced with degenerate or non-degenerate HNLs. However, if the precision in the determination of the mass splitting would be improved to reach the level of the large $\Delta M$ solutions, but only an upper bound is found, then this could indicate that 
a successful explanation of the baryon asymmetry requires degenerate neutrinos.



\section{Conclusion}
\label{sec:conclu}

We studied the minimal Type I seesaw model with two singlet fermions (HNLs) which are \textit{exactly} degenerate above the electroweak phase transition. This limit is interesting because the CP violation is fully encoded in the phases of the PMNS matrix.
In particular, we focused on the generation of the baryon asymmetry within the model for HNL masses below the Standard Model electroweak scale.
This region of parameter space is specially interesting because it can be tested in future experiments as SHiP, MATHUSLA and FCC-ee, as long as the model leads to HNL mixings beyond the naive see saw expectation $U^2 \gg m_\nu/M$.
Such mixings are achieved via an underlying lepton number symmetry, which we assume to be broken only by small Yukawa interactions.
This is commonly referred to as the linear seesaw limit~\cite{Akhmedov:1995ip,Akhmedov:1995vm}.
We can use this approximate symmetry to perturbatively solve for the baryon asymmetry.
The exact parameter dependence can be revealed by expressing the analytical solution in terms of CP invariants.
This allows to directly relate the baryon asymmetry to the neutrino parameters: the HNL mass and mixings as well as the CP violating phases of the PMNS matrix.

We find that the observed baryon asymmetry is only compatible with significant flavour hierarchies in the Yukawa interactions, which 
restrict the CP violating phases, see Figs.~\ref{fig:f_eps_fw_ov} and \ref{fig:scan_deltaphi}.
In particular, in the inverted hierarchy case we find that the maximal flavour hierarchy is achieved for CP conserving values of the Dirac and Majorana phases, where the baryon asymmetry vanishes.
This non-trivial interplay between the baryon asymmetry and the flavour structure of the model leads to a \textit{surprisingly} restrictive upper bound on the HNL mixing for which the observed baryon asymmetry can be explained, see Fig.~\ref{fig:scan}.
Similar conclusions also apply to the normal hierarchy scenario.

We further show that the baryon asymmetry can be predicted from laboratory measurements of the HNL mass and mixings, combined with an accurate determination of CP violation in neutrino oscillations. Additional constraints come from neutrinoless double beta decay searches. 
We explicitly demonstrate the prediction of the baryon asymmetry as well as the rate of neutrinoless double beta decay from a sequence of measurements of different flavour mixings and the Dirac phase, see Fig.~\ref{fig:reconstruction}.



\begin{acknowledgments}

We thank Claudia Hagedorn and Juraj Klari\'c for useful discussions. 
We also thank James M. Cline and Filipe Joaquim for useful feedback on the previous version of the manuscript. 
This work was partially supported by the EU H2020 research and innovation programme under the MSC grant agreement No 860881-HIDDeN, and the Staff Exchange grant agreement  No-101086085-ASYMMETRY, as well as by the Spanish Ministerio de Ciencia e Innovacion project PID2020-113644GB-I00, the Spanish Research Agency (Agencia Estatal de Investigaci\'on) through the project CNS2022-136013 and by Generalitat Valenciana through the ``plan GenT" program (CIDEGENT/2018/019) and grant PROMETEO/2019/083.
The work of SS received the support of a fellowship from ``la Caixa" Foundation (ID 100010434) with fellowship code LCF/BQ/DI19/11730034.
SS also thanks the CERN theory group, the Lawrence Berkeley National Laboratory and the Berkeley Center for Theoretical Physics for hospitality.

\end{acknowledgments}


\newpage
\appendix


\section{Appendix: Constraints from light neutrino data}
\label{app:param}

In this appendix we quote the structure of the Yukawa matrices for the normal and inverted hierarchy scenario when properly taking the light neutrino data constraints into account.
Recall that the model of Eq.~\eqref{eq:lag} posses an underlying lepton number symmetry of the form $L(N_1) = -L(N_2) = 1$ which leads to the most general form of the Yukawa and Majorana mass matrix
\be
\label{eq:app:param}
Y_{\alpha i}=\begin{pmatrix}
y_{e} e^{i \varphi_e} & y'_e e^{i \varphi_e'}\\
y_{\mu} e^{i \varphi_\mu} & y'_{\mu}  e^{i \varphi_\mu'}\\\
y_{\tau} e^{i \varphi_\tau} & y'_{\tau} e^{i \varphi_\tau'}\ 
\end{pmatrix},\;\;
M_R=\begin{pmatrix}
0 & \Lambda \\
\Lambda  & 0 
\end{pmatrix}\,.
\ee
When taking light neutrino oscillations data into account the structure of the Yukawa matrix gets tightly constraint.
Not all parameters are free and their correlation can be found via the seesaw relation
\be
- \left( m_{\nu}\right)_{\alpha\beta} = \frac{v^2}{\Lambda}\left( Y_{\alpha 1} Y_{\beta 2} + Y_{\alpha 2} Y_{\beta 1} \right)=\left(U^*m\,U^\dagger\right)_{\alpha\beta}\,,
\ee 
where $U=U(\theta_{12},\theta_{13},\theta_{23},\delta,\phi)$  is the PMNS matrix\footnote{We use the parameterization of the PDG \cite{ParticleDataGroup:2020ssz}.} describing the light neutrino mixing observed in neutrino oscillation experiments, and $m$ is the diagonal matrix of the light neutrino masses. 
Note that in this minimal model the lightest neutrino mass is zero and, consequently, only one of the two Majorana phases present in the PMNS matrix is physical. Similarly, since $\Delta M=0$, the extra phase associated to the heavy sector relevant in the non completely degenerate case becomes unphysical.\footnote{This is easy to check taking the $\Delta M=0$ limit in the expressions of the Yukawa couplings given in Sec.~3 of~\cite{Hernandez:2022ivz}.}
The general expressions of the Yukawa matrix as a function of the PMNS and neutrino mass parameters can be found e.g. in~\cite{Hernandez:2022ivz}.
The limit of $\Delta M \to 0$ is summarized in the following.
%
%
%
%
%
\vspace{0.25 cm}
\newline
\noindent
\textbf{Normal Hierarchy.}
The Yukawas satisfy
\bea
\label{eq:Yno_PMNS}
Y_{\alpha 1}&=&\frac{y}{\sqrt{2}}\left(U^*_{\alpha 3}\sqrt{1+\rho}+U^*_{\alpha 2}\sqrt{1-\rho}\right),\nonumber\\
Y_{\alpha 2}&=&\frac{y'}{\sqrt{2}}\left(U^*_{\alpha 3}\sqrt{1+\rho}-U^*_{\alpha 2}\sqrt{1-\rho}\right)\,,\nonumber\\
\eea
where $y$ is a real, free parameter and\footnote{In this parameterization $m_3<0$ ($m_2<0$) for NH (IH)~\cite{Gavela:2009cd}. This negative sign can be reabsorbed with a redefinition of the Majorana phase included in the PMNS matrix $U_\nu$.} 
\bea
\rho = \frac{\sqrt{\Delta m^2_{\rm atm}}-\sqrt{\Delta m^2_{\rm sol}}}{\sqrt{\Delta m^2_{\rm atm}}+\sqrt{\Delta m^2_{\rm sol}}},\;\;\;\;\; y' = \frac{M}{2v^2y}\left(\sqrt{\Delta m^2_{\rm atm}}+\sqrt{\Delta m^2_{\rm sol}}\right).
\label{rhoNH}
\eea
\vspace{0.25 cm}
\newline
\noindent
\textbf{Inverted Hierarchy.}
In this case, we have
\bea
\label{eq:Yio_PMNS}
Y_{\alpha 1}&=&\frac{y}{\sqrt{2}}\left(U^*_{\alpha 2}\sqrt{1+\rho}+U^*_{\alpha 1}\sqrt{1-\rho}\right),\nonumber\\
Y_{\alpha 2}&=&\frac{y'}{\sqrt{2}}\left(U^*_{\alpha 2}\sqrt{1+\rho}-U^*_{\alpha 1}\sqrt{1-\rho}\right)\,,\nonumber\\
\eea
where, again, $y$ is real and arbitrary, while
\begin{align}
\label{rhoIH}
\begin{split}
\rho &= \frac{\sqrt{\Delta m^2_{\rm atm}}-\sqrt{\Delta m^2_{\rm atm}-\Delta m^2_{\rm sol}}}{\sqrt{\Delta m^2_{\rm atm}}+\sqrt{\Delta m^2_{\rm atm}-\Delta m^2_{\rm sol}}},\\
y' &= \frac{M}{2v^2y}\left(\sqrt{\Delta m^2_{\rm atm}}+\sqrt{\Delta m^2_{\rm atm}-\Delta m^2_{\rm sol}}\right)\,.
\end{split}
\end{align}



\section{Appendix: Leptogenesis with strong flavour washout}
\label{app:unflavoured}

We devote this appendix to derive the fact that the minimal model we discuss, see Sec.~\ref{sec:model}, can not accommodate the observed baryon asymmetry when all flavours are in strong washout, independently of any other appearing weak modes
\be
\label{eq:gamma_unflavoured_wHC}
\Gamma_{\rm LN}(T_{\rm EW}), \Gamma_{M}(T_{\rm EW}) < H_u(T_{\rm EW}) <  \Gamma_\alpha(T_{\rm EW}),\Gamma(T_{\rm EW})\,,
\ee
or
\be
\label{eq:gamma_unflavoured_sHC}
\Gamma_{\rm LN}(T_{\rm EW}) < H_u(T_{\rm EW}) <   \Gamma_{M}(T_{\rm EW}),\Gamma_\alpha(T_{\rm EW}),\Gamma(T_{\rm EW})\,,
\ee
see also the discussion in Sec.~\ref{subsec:sakharov}.
Using the flavour thermalization rates found in Eq.~\eqref{eq:Gamma_alpha}, and taking light neutrino constraints into account, see Eq.~\eqref{eq:mnuU}, the mixing for which all flavours are in strong washout corresponds to
\be
\label{eq:U2_regime_2}
 U^2 \geq 10^{-9} \left({1{\rm GeV}\over M}\right)^2{1\over {\rm Min}(\epsilon_\alpha)}\,,
\ee
with
\begin{eqnarray}
\label{eq:app:minepsalpha}
{\rm Min}(\epsilon_e)_{NH} \simeq  {\rm Min}(\epsilon_\tau)_{IH} \simeq 10 \times {\rm Min}(\epsilon_\mu)_{IH} \simeq 5\times 10^{-3}  \,.
\end{eqnarray}
This corresponds to the blue region above the red dashed line in Fig.~\ref{fig:regimes}.
We call this region the \textit{unflavoured} regime.
It can be further divided into two regions depending whether helicity conserving rates are weak, corresponding to Eq.~\eqref{eq:gamma_unflavoured_wHC}, or strong, corresponding to Eq.~\eqref{eq:gamma_unflavoured_sHC}, at the EPWT.
This corresponds to the region below and above the blue dashed line in Fig.~\ref{fig:regimes}, respectively.  
In the unflavoured regimes with wHC the asymmetry is expected to receive contributions of the form
\begin{align}
\Delta^M \equiv \frac{1}{\rm{Tr}\left( Y^\dagger Y \right)^2}  \sum_{\alpha} \frac{\Delta_\alpha}{\left( Y Y^\dagger \right)_{\alpha\alpha}}  \,.
\label{eq:cpinv_unflavoured_weak}
\end{align}
We will show in the following it leads to insufficient baryon asymmetry. 
On the other hand, in the unflavoured sHC regime, these invariants do not contribute and the relevant ones start at ${\mathcal O}(y'^4)$ and therefore can be neglected at the order we work.\footnote{We have also checked numerically that the observed baryon asymmetry can not be generated in this regime.}

In terms of the parametrization of Eq.~\eqref{eq:LNVparam2N} (equivalently Eq.~\eqref{eq:app:param}) the CP invariant of Eq.~\eqref{eq:cpinv_unflavoured_weak} can be expressed as
\be
\label{eq:cpinv_unflavoured_param}
\Delta^{\rm M} =  \frac{2 M^2 }{y^2} \sum_{\alpha < \beta} (y_\alpha^2 - y_\beta^2) \frac{y'_\alpha y'_\beta}{y_\alpha y_\beta} \sin(\Delta \varphi_\alpha - \Delta\varphi_\beta)\,,
\ee
with $\Delta \varphi = \varphi' - \varphi$.
When taking the constraints from light neutrino data into account, the invariant can be expressed in terms of the observables $(M, |U^2|, \delta, \phi)$, see Sec.~\ref{sec:cpinv}.
The result at leading order in $y'/y$ as well as leading order in the small light neutrino parameters 
\begin{align}
r \equiv \frac{\sqrt{\Delta m_{\rm sol}^2}}{\sqrt{\Delta m_{\rm atm}^2}} \sim \theta_{13} \sim |\theta_{23} - \pi/4| \sim 10^{-1}\,,
\end{align}
is given by
\bea
\label{eq:app:CP_inv_U2_M_M_NH}
\Delta^{\rm M} &=&  - \frac {\Delta m_{\rm{atm}}^2 }{4 U^4\sqrt{r} } \frac{ \theta_{13} \sin(\delta + \phi)}{s_{12}} \,\,\, (\rm{NH}) \,, \\
\label{eq:app:CP_inv_U2_M_M_IH}
 \Delta^{\rm M} &=& \frac{\Delta m_{\rm{atm}}^2}{4 U^4}  \frac{r^2 c_{12} s_{12} \sin\phi \left( 1 + 6 c_{12} s_{12} \cos\phi \right)}{4 c_{12}^2 s_{12}^2 \cos^2\phi - 1} \,\,\, (\rm{IH}) \,.
\eea
Following the perturbative expansion as outlined in Sec.~\ref{subsec:pert_approach} to solve the quantum kinetic equations of Eq.~\eqref{eq:rhonrhonbarav} we can derive
\begin{align}
\label{eq:asym_ov_wlnv}
\sum_{\alpha } \mu_{B/3 - L_\alpha} = - x^2 \frac{8 \gamma_0 \kappa (s_0 \omega + \gamma_0 \omega_M) }{(6 \gamma_0 + \gamma_1 \kappa) (\gamma_0^2 + 4 \omega^2)} \frac{ \Delta^{\mathrm{M}} }{T_{\mathrm{EW}}^2}\,.
\end{align}
Evaluating the interaction rates at $T=10^6\,\mathrm{GeV}$, see Tab.~\ref{tab:rates}, and for masses $M=1\,\mathrm{GeV}$, this can be expressed explicitly as 
\be
Y_B = 4.3\times10^{-26} \left( \frac{1}{U^2} \right)^2 f_{\rm M}^{\rm H}\,,
\ee
where the function $f_{\rm M}^{\rm H}$ encodes the angular dependence of the CP invariants.
They can be obtained from Eq.~\eqref{eq:app:CP_inv_U2_M_M_NH} (Eq.~\eqref{eq:app:CP_inv_U2_M_M_IH}) for NH (IH):
\begin{align}
\label{eq:f_M_NH}
f_{\rm M}^{\rm NH} =  \frac{\theta_{13} \sin(\delta + \phi)}{\sqrt{r} s_{12}}\,, \,\,\,
f_{\rm M}^{\rm IH} = -  \frac{r^2 c_{12} s_{12} \sin\phi \left( 1 + 6 c_{12} s_{12} \cos\phi \right)}{4 c_{12}^2 s_{12}^2 \cos^2\phi - 1} \,. 
\end{align}
Maximizing the angular functions and demanding to match the observed baryon asymmetry leads to
\begin{align}
|U^2| \lesssim 18\,(6) \times 10^{-9} \, \rm{NH\, (IH)}\,,
\end{align}
which is always smaller than the required mixing of Eq.~\eqref{eq:U2_regime_2} for the relevant masses.
Therefore, the observed baryon asymmetry can not be explained within this regime.

This also means that the observed asymmetry in the unflavoured sHC regime can not be explained, indeed confirming the expectation that the relevant $\mathcal{O}(y'^4)$ CP invariant is too small.



\bibliography{biblio}

\end{document}